\documentclass[prc,preprintnumbers,unsortedaddress,amsmath,amssymb,floatfix]{revtex4}
\usepackage{comment}
\usepackage{amssymb}
\usepackage{color}
\usepackage{rotating}
\usepackage{amsmath}
\usepackage{ulem}
\usepackage{overpic}
\usepackage{graphicx}
\usepackage{dcolumn}
\usepackage{graphicx}
\usepackage{bm}
\usepackage{chngpage}
\usepackage{multirow}
\usepackage{slashed}
\usepackage{indentfirst}
\usepackage{booktabs}
\usepackage{amssymb,amsfonts}
\usepackage{amsmath}
\usepackage{xcolor}
\usepackage[colorlinks,citecolor=blue,linkcolor=blue,hypertex,breaklinks=true]{hyperref}

\newcommand{\be}{\begin{equation}}
\newcommand{\ee}{\end{equation}}
\newcommand{\bea}{\begin{eqnarray}}
\newcommand{\eea}{\end{eqnarray}}

\begin{document}
\title{Microscopic Optical Potential from Brueckner-Hartree-Fock Theory}

\author {Miao Qi$^{1}$, Li-Li Chen$^{1}$, Li-Gang Cao$^{1,2}$\footnote{Corresponding author: caolg@bnu.edu.cn}, Feng-Shou Zhang$^{1,2}$, Xin-Le Shang$^{3,4}$, Wei Zuo$^{3,4}$, and U. Lombardo$^{5}$ }

\affiliation{
$^{1}$Key Laboratory of Beam Technology of Ministry of Education,\\
School of Physics and Astronomy, Beijing Normal University, Beijing 100875, China\\
$^{2}$Key Laboratory of Beam Technology of Ministry of Education, Institute of Radiation Technology,\\
Beijing Academy of Science and Technology, Beijing 100875, China\\
$^{3}$CAS Key Laboratory of High Precision Nuclear Spectroscopy, Institute of Modern Physics,
Chinese Academy of Sciences, Lanzhou 730000, China\\
$^{4}$School of Nuclear Science and Technology, University of Chinese Academy of Sciences, Beijing 100049, China\\
$^{5}$Dipartimento di Fisica and Laboratori Nazionali del Sud (INFN), Via S. Sofia 64, I-95123 Catania, Italy
 }

\date{\today}

\begin{abstract}
Modern Brueckner-Hartree-Fock (BHF) calculations are very successful in describing various properties of symmetric and asymmetric nuclear matter. Within BHF theory a microscopic optical potential (MOP) for nucleon-nucleus scattering is developed. First, we parametrize the energy and density dependence of complex optical potentials in nuclear matter based on BHF calculations and then we construct the MOP for finite nuclei with the local density approximation extended to
include the finite-range effects. The density
distribution and the spin-orbit contribution are calculated from the Hartree-Fock (HF) approximation with LNS5 Skyrme interaction, the latter being constrained by the BHF results. The central real and imaginary potentials turn out to be quantitatively consistent with
the phenomenological global Koning-Delaroche (KD) potentials.  The performance of MOP is evaluated by considering neutron/proton scattering on $^{40,48}$Ca. The elastic scattering differential cross sections, analyzing powers and total/reaction cross sections are analyzed
in the energy below 200 MeV. A good agreement between the theoretical results and the measurements is achieved. Since our results are presented in the analytic forms, they can thus be used easily in the analysis of the experimental data of the nucleon scattering on exotic nuclei.
\end{abstract}

\maketitle

\section{Introduction}\label{sec1}

Nuclear reactions provide an effective tool to study the structure and the dynamics of atomic
nuclei as well as the properties of nuclear force in the laboratory~\cite{Tanihata95,Johnson20}.  They also play crucial role in nuclear astrophysics for the understanding of the evolution of stars and the origin of elements in the cosmos~\cite{Kasen17,Cowan21,Schatz08}. In particular, the nuclear reactions are
the main goal of many experimental programs
at rare-isotopes beam facilities worldwide. To analyse the nuclear reaction experimental data, the optical model is one of the most fundamental theoretical
tools, in which an effective complex and energy dependent potential is usually adopted to describe the interaction between the projectile and
target~\cite{Feshbach58,Hodgson63}. The optical potential is characterized by a real part
describing the elastic channel, and an imaginary part which
takes into account the flux loss from the elastic
channel into other channels.

The optical potential can be obtained either phenomenologically
or microscopically. In general, the various contributions to the phenomenological
optical potential are assumed to have an analytic form, such as the Wood-Saxon type, which customarily
is used also to describe the nuclear density profile. It depends on
adjustable parameters that are functions of the energy and
nuclear mass number~\cite{Varner91,Koning03}. These parameters are adjusted to optimize
the fit to experimental data of elastic scattering on stable nuclei. But, the application of these potentials to study nuclei far from the stability turns out to be unreliable with uncontrolled uncertainties.
With the recent development of rare-isotopes beams, the nuclei far from stability are now becoming accessible. Therefore, more effort is requested to study  in the near future the systems far from stability. To enter the field of nuclear reactions with exotic nuclei, it is critical to connect the optical potential to an underlying microscopic theory, which can guarantee reliable predictions either for stable nuclei  and nuclei far from the stability.

It is well known that the microscopic optical potential (MOP) for scattering states is identified with the on-shell
energy and momentum dependent single-particle self-energy in quantum many-body theory~\cite{bell59}. This identification makes it possible to utilize the
many-body technique to obtain the MOP without free parameters. Great efforts have been made to develop the MOP from realistic nucleon-nucleon(NN) interactions, such as Nijmegen~\cite{nij}, Paris~\cite{paris}, Bonn~\cite{bonn}, Argonne~\cite{argo}, and the interactions developed more recently from chiral effective-field theory~\cite{van,epe,mach}.
Different many-body approaches have been taken to build MOP in recent years, for instance the self-consistent
Green's function approach, the inversion
of a Green's function based on coupled-cluster or
on no-core shell model calculations, chiral
symmetry inspired optical potential, double-folding potentials
from chiral effective field theory, Brueckner-Hartree-Fock (BHF) $G$-matrix calculations, and dispersive optical potential ~\cite{Whitehead21,Whitehead19,Holt13,Vorabbi22,Vorabbi24,Durant22,Rotureau18,Idini19,Burrows24,Nollett07,Hupin13,Birse20,Kohno20,Qin24,xu12,xu16,jlmapp19,Atkinson20}. Recent review papers on MOP can be found in Refs.~\cite{Dickhoff19,Rotureau20,Holt22,Hebborn23,Finelli23}. The methods, which are applied directly to build the MOP for finite nucleus, work quite well for the light nuclei. But, with upcoming experiments on exotic nuclei in the medium and heavy-mass regions of the nuclear chart, they prove inadequate so that much effort is demanded to develop microscopic reaction theories that start from realistic nucleon-nucleon interaction. This is the main task of the nuclear reaction community for the next decade.

The nuclear matter approach is a successful method in the study of MOP. The framework to employ nuclear matter calculations of MOP in finite nuclei was built in early pioneering
works~\cite{Jeuke76,Jeuke77,Brieva771,Brieva772}, since infinite homogeneous nuclear matter is a relatively simpler physical system compared to finite nuclei. This, in fact, enables us to set a general form of the self-energy and, correspondingly, of a nucleon optical potential, which can cover a relative large range of density and isospin asymmetry, so to avoid many of the technical difficulties and practical limitations faced when computing the nucleon optical potential directly in finite nucleus.
Within the local density approximation(LDA), nucleon-nucleus optical potential can be obtained by folding the nuclear matter optical potential with the isoscalar and isovector densities of the target nuclei. The optical potential obtained  can  be applied in wide regions of the nuclear chart and it will be vital for the future of nuclear reactions of exotic nuclei.

Recently, the microscopic BHF theory has been developed  to a large extend and it has achieved great success in describing the properties of nuclear matter. Now, the calculations can be consistently extended to the asymmetric nuclear matter~\cite{Bomb91,Zuo99,Baldo88} with  the inclusion of three-body forces(TBF)~\cite{Zuo02,lizh081,Zuo06}. Very important results have been obtained in the description of equation of state of nuclear matter, including the incompressibility and symmetry energy, and the critical endpoint of the liquid-to-gas phase transition~\cite{cao06,Gamba11}.
The BHF theory has also been successfully applied to study proton and neutron single-particle potentials, effective masses, and  isospin dependence in neutron rich nuclear matter as well as neutron star properties~\cite{Zuo05,Zuo04,shang20,lizh22}.
The MOP based on BHF approach with a new version of three-body
force is performed to interpret the volume part of the phenomenological optical potential~\cite{lill09,lizh08}, but the calculations have not yet been extended to the optical potentials of finite nuclei. In Ref.~\cite{Shang21}, an exact  treatment of the total momentum is adopted in solving the nuclear matter
Bruecker-Bethe-Goldstone(BBG) equation with two and three body forces. The single-particle potential, the equation of
state, and the nucleon effective mass can be calculated consistently from the $G$-matrix~\cite{Shang22,Shang25}. So we think that it is time to build the MOP for finite nuclei from the microscopic BHF theory with realistic nucleon-nucleon interactions.

In this paper, the MOP is obtained first in symmetric and asymmetric nuclear matter, then the MOP for finite nucleus will be deduced with the improved local density approximation(ILDA).
To test our MOP, the elastic scattering differential cross sections, analyzing powers and total/reaction cross sections for neutron/proton(n/p)~+~$^{40,48}$Ca elastic scattering are
investigated.

This paper is organized as follows. The theoretical formalism
of the BHF theory and the MOP are introduced in
Sec. II. In Sec. III, first, the MOP in nuclear matter are parametrized using the ansatz of the Jeukenne-Lejeune-Mahaux~\cite{Jeuke77}. We then calculate the nuclear density distribution from Skyrme Hartree-Fock theory and employ
the ILDA to construct nucleon-nucleus optical potentials
for n/p~+~$^{40,48}$Ca reactions. We compute the elastic scattering differential cross sections, analyzing powers and total/reaction cross sections. These results are compared with the experimental data and the predictions from the global Koning-Delaroche (KD) phenomenological
optical potential~\cite{Koning03}.
Finally, in Sec. IV, the overall presentation
is summarized and some prospects are given.

\section{Theoretical framework}\label{sec2}
\subsection{Bruecker-Bethe-Goldstone equation}

The Brueckner theory for symmetric and asymmetric nuclear matter and its extension to include TBF are
described in details in Refs.~\cite{Bomb91,Zuo99,Baldo88,Zuo02,lizh081,Zuo06,Shang21}. Here we simply give a brief review
for completeness. The starting point of the BHF approach is
the reaction $G$-matrix, which satisfies the isospin-dependent Brueckner-Bethe-Goldstone equation,

\begin{eqnarray}\label{eq:G Marix}
		G(\rho,\beta,\omega)=v_{NN}
		+v_{NN}\sum_{k_1k_2}\frac{\left|k_1k_2\right\rangle Q(k_1,k_2)\left\langle{k_1k_2}\right|}{{\omega}-\boldsymbol{\varepsilon}(k_1)-{\varepsilon}(k_2)+i{\eta}}G(\rho,\beta,\omega),
\end{eqnarray}
where $v_{NN}$ is the realistic nucleon-nucleon interaction, $k_i\equiv(\vec k_i,\sigma_i,\tau_i)$ denotes the single
particle momentum, the $z$-component of spin and isospin,
respectively. $Q$, $\omega$ and $\varepsilon(k)$ are the Pauli operator, the starting energy, and the single-particle energy, respectively. $\beta=(\rho_n-\rho_p)/\rho$ is the asymmetry parameter, where $\rho,~\rho_n$, and $\rho_p$ denote the total, neutron and proton
densities in nuclear matter.  When solving the BBG equation, the continuous
choice for the auxiliary potential is usually adopted since
it provides a much faster convergence of the hole-line expansion
than the gap choice~\cite{Jeuke76,song98}. An advantage of the
continuous choice is that the auxiliary potential has the physical
meaning of the mean field felt by a nucleon during its propagation
between two successive scatterings in nuclear
medium~\cite{sartor99}.

To avoid the difficulties of the full treatment of the three-body problem
(for a detailed description and justification of the method we refer to Ref.~\cite{grange89}), the microscopic TBF is approximated by an effective two-body force, $\it integrating over$ the third nucleon degrees of freedom as it follows
\begin{equation}\label{eq:TBF}
\begin{aligned}
\langle\vec{r}_1\vec{r}_2|V_3|\vec{r}_1^{\prime}\vec{r}_2^{\prime}\rangle & =\frac{1}{4}\operatorname{Tr}\sum_n\int\mathrm{d}\vec{r}_3\mathrm{d}\vec{r}_3^{\prime}\phi_n^*\left(\vec{r}_3^{\prime}\right)
\left(1-\eta\left(r_{13}^{\prime}\right)\right)\left(1-\eta\left(r_{23}^{\prime}\right)\right) \\
 & \times W_3\left(\vec{r}_1^{\prime}\vec{r}_2^{\prime}\vec{r}_3^{\prime}|\vec{r}_1\vec{r}_2\vec{r}_3\right)\phi_n(r_3)\left(1-\eta(r_{13})
 \right)\left(1-\eta(r_{23})\right),
\end{aligned}\end{equation}
where the effect of correlations between the third particle and the two others is weighted by the defect function $\eta(r)$ (see Ref.~\cite{grange89}).
Because of its dependence  on the defect function, the effective two-body force is recalculated self-consistently along with the $G$-matrix at each
step of the iterative BHF procedure.

\subsection{Nucleon Self-energy and microscopic optical potential}

In the BHF approximation with TBF, the self-energy contains three main contributions,
\begin{equation}
\Sigma_\tau(k,E)=\Sigma_\tau^{\mathrm{bhf}}(k,E)+\Sigma_\tau^{\mathrm{cpol}}(k,E)
+\Sigma_\tau^{\mathrm{tbf}}(k).
\end{equation}
The self-energy, in general, is a complex quantity that is
energy and momentum dependent.
The first term is the BHF mean field potential with the G-matrix as
the effective interaction, the second term is due to the core
polarization, and the third term, called TBF rearrangement
term, is due to the density dependence
of the effective three-body force. The first two terms from the BHF approach have been discussed in Refs.~\cite{Zuo02,Zuo05} in the case of
asymmetric nuclear matter.
In general, the TBF effect  is twofold: first, it affects the self-energy
via the modification of the G-matrix. This effect has been
embodied in the BHF self-energy, i.e., the first two terms in
Eq. (3); second, the density dependence of TBF introduces  an
additional contribution, i.e., a rearrangement contribution to
the self-energy (third term in Eq. (3))~\cite{Zuo06}. The last term, in our approximation is real, being
directly related to the TBF.

In nuclear matter, the nuclear optical potential is given by the on-shell matrix elements of the self-energy $\Sigma_\tau(k(E),E)= \Sigma_\tau(E)$, $k(E)$ being obtained from the mass shell relation.
The MOP for a nucleon in nuclear matter depends on energy $E$, density $\rho$, and isospin asymmetry $\beta$. The dependence of the
optical potential on the isospin asymmetry $\beta$ is crucial for scattering on exotic nuclei. The Lane empirical potential~\cite{lane62,bauge01,Holt16,bauge98} is widely used in
phenomenological and, surprisingly, confirmed by nuclear-matter microscopic optical potentials in a wide range of symmetries~\cite{Bomb91}.
So, in asymmetric nuclear matter, the MOP can be assumed to be linear in the
asymmetry parameter and split into the
isoscalar and isovector components,
\begin{equation}
U_{NM}(\rho,E,\beta)=V_0(\rho,E)+\tau \beta  V_1(\rho,E)+i\frac{m^*_{k,\tau}}{m}[W_{0}(\rho,E)+\tau\beta W_{1}(\rho,E)],
\end{equation}
where $V_0 (W_{0})$ is the depth of the isoscalar potential, while $V_1 (W_{1})$ is the depth of the isovector potential. $\tau=-1$ is for proton and $\tau=1$ is for neutron projectiles. $m^*_{k,\tau}/m$ is the $k$-mass of neutron or proton. In the framework of the microscopic approach, the effective k-mass embodies the nonlocality of the optical potential ~\cite{Jeuke76,Jeuke77}, that is in keeping with the approach addressed
in Refs. \cite{Arell24,Arell22}. $V_0$ and $W_0$ are potential components in symmetric nuclear matter defined as the follows
\begin{equation}
V_0(\rho,E)= \text{Re} \Sigma_\tau(k,E,\beta=0),
\end{equation}
\begin{equation}
W_0(\rho,E)= \text{Im} \Sigma_\tau(k,E,\beta=0).
\end{equation}
The symmetry potentials $V_{1}$ and $W_{1}$ have the following expressions\cite{lill09,lizh08},
\begin{equation}
V_{1}(\rho,E)=\frac{\text{Re}[\Sigma_{n}(\rho,E,\beta)-\Sigma_{p}(\rho,E,\beta)]}{2\beta},
\end{equation}
\begin{equation}
W_{1}(\rho,E)=\frac{\text{Im}[\Sigma_{n}(\rho,E,\beta)-\Sigma_{p}(\rho,E,\beta)]}{2\beta}.
\end{equation}

In nuclear matter there is no Coulomb force between
protons, but for the proton induced nuclear reactions with finite nuclei, one must introduce the Coulomb correction into
the proton optical potential to get agreement with experimental data. This is given by the
following prescription~\cite{Jeuke77,Jeuke7715},
\begin{equation}
\Sigma_{p}^c=\Sigma_{p}(E-V_c)+V_c,
\end{equation}
where $V_c$ denotes the Coulomb potential.

\section{Results and Discussion}\label{sec3}

\subsection{Parameterization of the MOP}

The binding energy per nucleon, the symmetry energy, the isospin dependence of the single-particle properties, the mean field, the effective mass, and
the mean free path of neutrons and protons from BHF calculations have been discussed in details in Refs.~\cite{Zuo99,Zuo05,lill09,lizh08,Shang21}. In this subsection, we focus on the parameterization of the MOP in nuclear matter. First we calculate the nucleon self-energies in different densities and asymmetric parameters with solving the Brueckner-Bethe-Goldstone equation~\cite{Shang21}, the realistic nucleon-nucleon Argonne V18 interaction and a consistent microscopic three-body interaction~\cite{argo,grange89,Zuo02} are adopted in our present calculations. We consider the energy domain 10 $\leq E \leq$ 200 MeV, and for the densities we consider $\rho =0.2, ~0.18, ~0.16, ~0.14, ~0.12, ~0.10, ~0.08$, and~ 0.06~fm$^{-3}$, respectively. The range of densities covers not only the nuclear core, but also part of the surface. To construct the MOP for finite nucleus, the self-energies in symmetric and asymmetric nuclear matter at lower densities are needed. However, it is well known
that the BHF calculations are not reliable for low density due to the cluster effects~\cite{Lombardo05,Arell15}. An extrapolation method is usually adopted to get the potentials for low densities with the natural constraint that the
potentials vanish at $\rho$ = 0~\cite{Qin24,xu16}. As mentioned in the introduction, we will use the Jeukenne-Lejeune-Mahaux ansatz\cite{Jeuke77} to fit the BHF results, which can guarantee the natural constraint at $\rho$ = 0.

\begin{figure}[htb]
   \centering
	\begin{minipage}{0.49\linewidth}
		\centerline{\includegraphics[width=7.0cm]{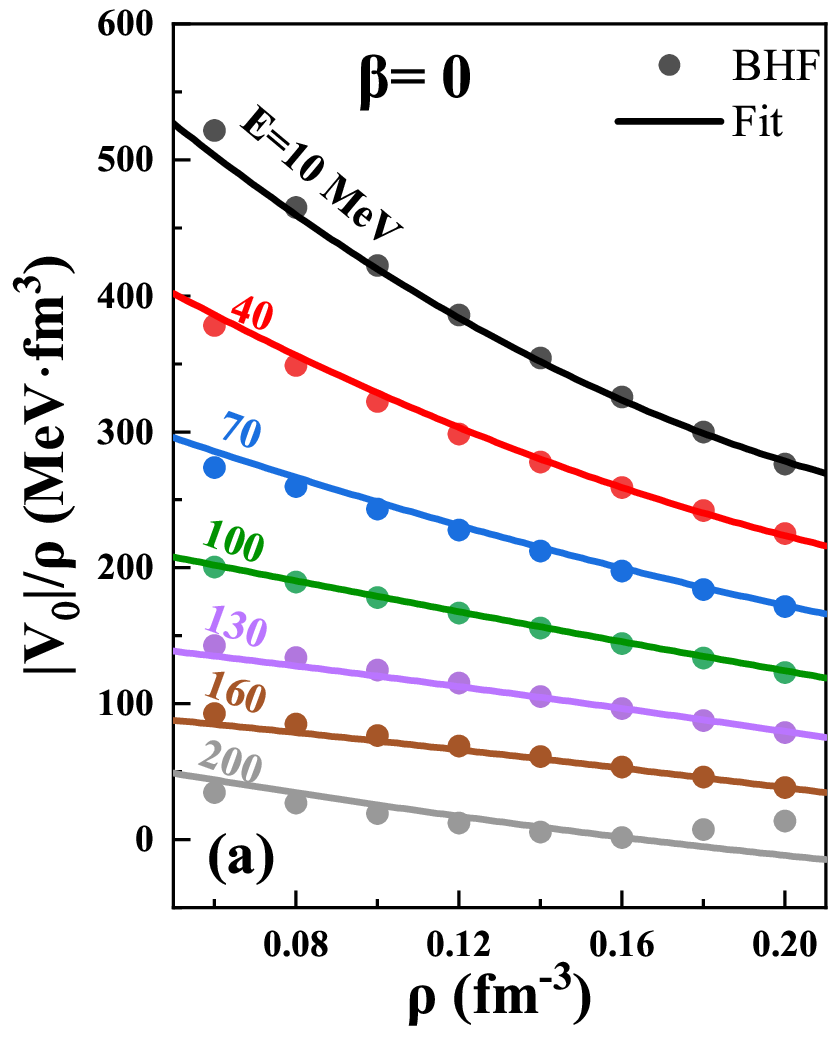}}
	\end{minipage}
	\begin{minipage}{0.49\linewidth}
		\centerline{\includegraphics[width=7.0cm]{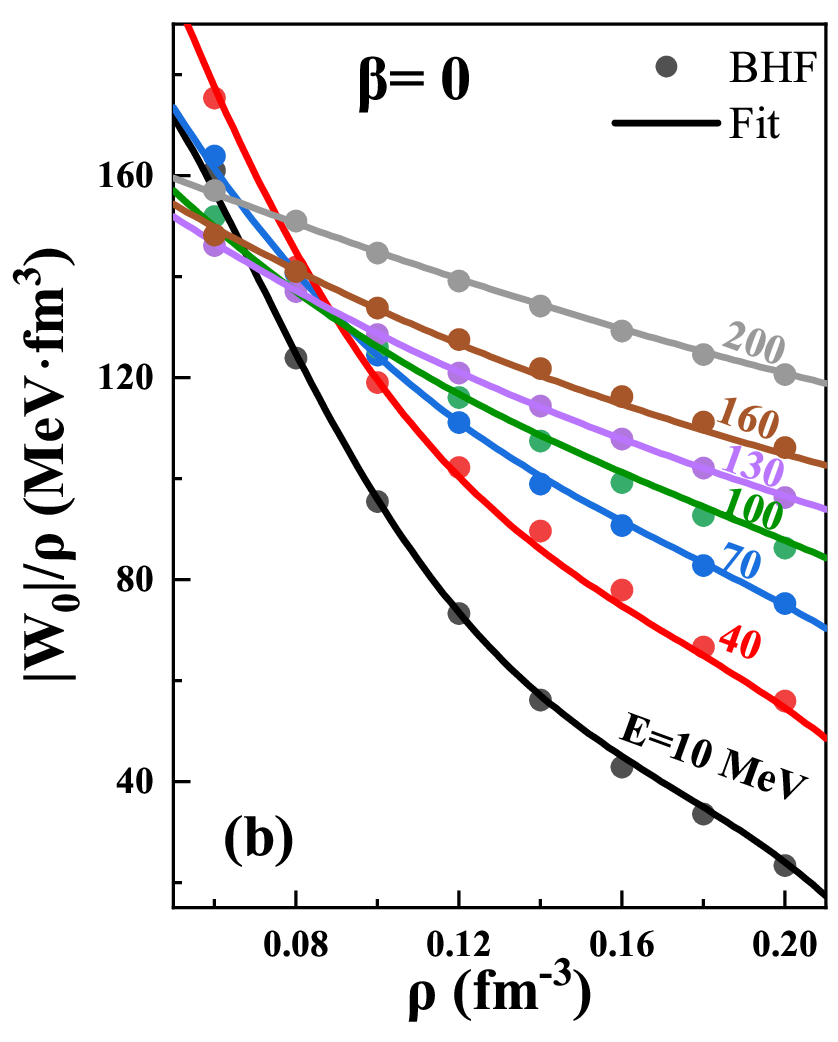}}
	\end{minipage}
	\caption{(Color online) The solid dots represent the dependence on the density of the quantities ~$|{V_0(\rho,E)}|/\rho$ (a) and~${|W_0(\rho,E)|/\rho}$ (b) for energies $E$=10, 40, 70, 100, 130, 160, and 200 MeV by BHF calculations. The Solid lines correspond to the parametrizations.}\label{VW0}
\end{figure}

In order to simplify the use of our MOP, we parametrized the BHF results in nuclear matter in the analytical forms from Refs.~\cite{Jeuke77,bauge98}.
As discussed in Refs.~\cite{Jeuke77,bauge98}, the quantity $V_0(\rho,E)$ is intimately related to the potential energy density which plays a central role
in the energy density approximation for finite nuclei, and the ratio $V_0/\rho$ is closely connected with the strength of the effective interaction used in the Thomas-Fermi approximation. In Fig. \ref{VW0}, we plot the ratio $|V_0(\rho,E)|/\rho$ (a) and $|W_0(\rho,E)|/\rho$ (b) versus $\rho$ for the energies at 10, 40, 70, 100, 130, 160, and 200 MeV, the solid dots are the results from BHF calculations, the lines are the fitted results. For the isoscalar part $V_0$ of real MOP, we have parametrized it in the form
\begin{equation}\label{eq:V0}
	V_0(\rho,E)=\sum_{i,j=1}^3a_{ij}\rho^iE^{j-1},
\end{equation}
where $E$ is expressed in MeV, $\rho$ in $\mathrm{fm}^{-3}$, and $a_{ij}$ are the coefficients given in Table \ref{tab:aij-Vo}. It is suggested in Refs.~\cite{Jeuke77,bauge98} that the parametric form of the isoscalar component $W_0(\rho,E)$ of the imaginary MOP should take into account the fact that it is proportional to $(E-\varepsilon_F)^2$ near the Fermi surface. A best fit to BHF results can be obtained with the expression
\begin{equation}\label{eq:W0}
W_0(\rho,E)=\left[1+\frac{D}{\left(E-\varepsilon_F\right)^2}\right]^{-1} \sum_{i,j=1}^4b_{ij}\rho^iE^{j-1},
\end{equation}
where $\varepsilon_F$ is the Fermi energy, and $D=600$~MeV$^2$ taken from Refs.\cite{Jeuke77,bauge98}, and $b_{ij}$ are the coefficients given in Table \ref{tab:bij-Wo}. For $\varepsilon_F$, we have calculated it using BHF approximation and fitted it with the expression
\begin{equation}
\varepsilon_{F}(\rho)=\rho(-487+2977\rho-6436\rho^2).
\end{equation}
From Fig. \ref{VW0}(a), it can be observed that the ratio $|V_0(\rho,E)|/\rho$ decreases with increasing energy and density. $|W_0(\rho,E)|/\rho$ always decreases with increasing density. In contrast, the behavior of $|W_0(\rho,E)|/\rho$ with energy looks quite interesting for $\rho<0.10\mathrm{fm}^{-3}$ as shown in Fig. \ref{VW0}(b), it increases from 10 to 40 MeV, then decreases until to 130 MeV, after that it increases again. At larger densities, the quantity $|W_0(\rho,E)|/\rho$ steadily increases with energy.

\begin{table}[htb]
	\centering
	\caption{Coefficients ~$a_{ij}$ ~in the expression~(\ref{eq:V0})~for~$V_0(\rho,E)$~.}\label{tab:aij-Vo}
	\setlength\tabcolsep{5mm}{
    \renewcommand{\arraystretch}{1.5}
    \begin{tabular}{cccc}\\
    \hline\hline
		{$a_{ij}$} & 1    & 2     & 3          \\
    \hline
		1 & $-7.2031 \times 10^{2}$ & $6.4239 \times 10^{0}$  & $-1.5952\times 10^{-2}$ \\
		2 & $3.2404 \times 10^{3}$  & $-3.9186 \times 10^{1}$ & $1.2890 \times 10^{-1}$ \\
		3 & $-5.6246 \times 10^{3}$ & $8.3259 \times 10^{1}$  & $-2.9125 \times 10^{-1}$\\
    \hline\hline
	\end{tabular}}
\end{table}

\begin{table}[htb]
	\centering
	\caption{Coefficients ~$b_{ij}$ ~in the expression~(\ref{eq:W0})~for~$W_0(\rho,E)$~.}
	\label{tab:bij-Wo}
	\setlength\tabcolsep{5mm}{
    \renewcommand{\arraystretch}{1.5}
		\begin{tabular}{ccccc}\\
			\hline\hline
      { $b_{ij}$} & 1    & 2     & 3    & 4   \\
			\hline
		1   & $-7.1887 \times 10^{2}$ & $9.9479 \times 10^{0}$  & $-6.1646\times 10^{-2}$ & $1.2710 \times 10^{-4}$  \\
		2   & $1.0004 \times 10^{4}$  & $-1.7376 \times 10^{2}$ & $1.0870 \times 10^{0}$  & $-2.2971 \times 10^{-3}$ \\
		3   & $-5.2545 \times 10^{4}$ & $9.4321 \times 10^{2}$  & $-5.8571 \times 10^{0}$ & $1.2225 \times 10^{-2}$  \\
	    4   & $9.9892 \times 10^{4}$  & $-1.7776 \times 10^{3}$ & $1.0813 \times 10^{1}$  & $-2.2107 \times 10^{-2}$ \\
			\hline\hline
	\end{tabular}}
\end{table}

\begin{figure}[htb]
	\begin{minipage}{0.49\linewidth}
		\centerline{\includegraphics[width=7.0cm]{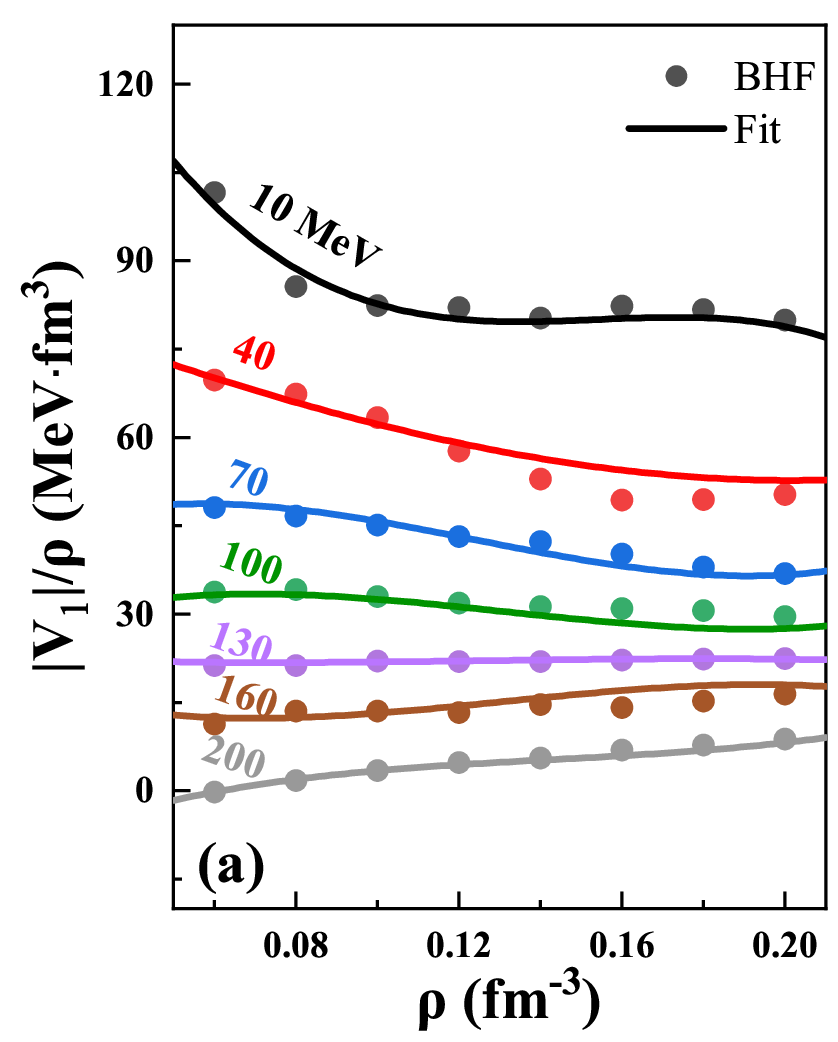}}
	\end{minipage}
    \hfill
	\begin{minipage}{0.49\linewidth}
		\centerline{\includegraphics[width=7.0cm]{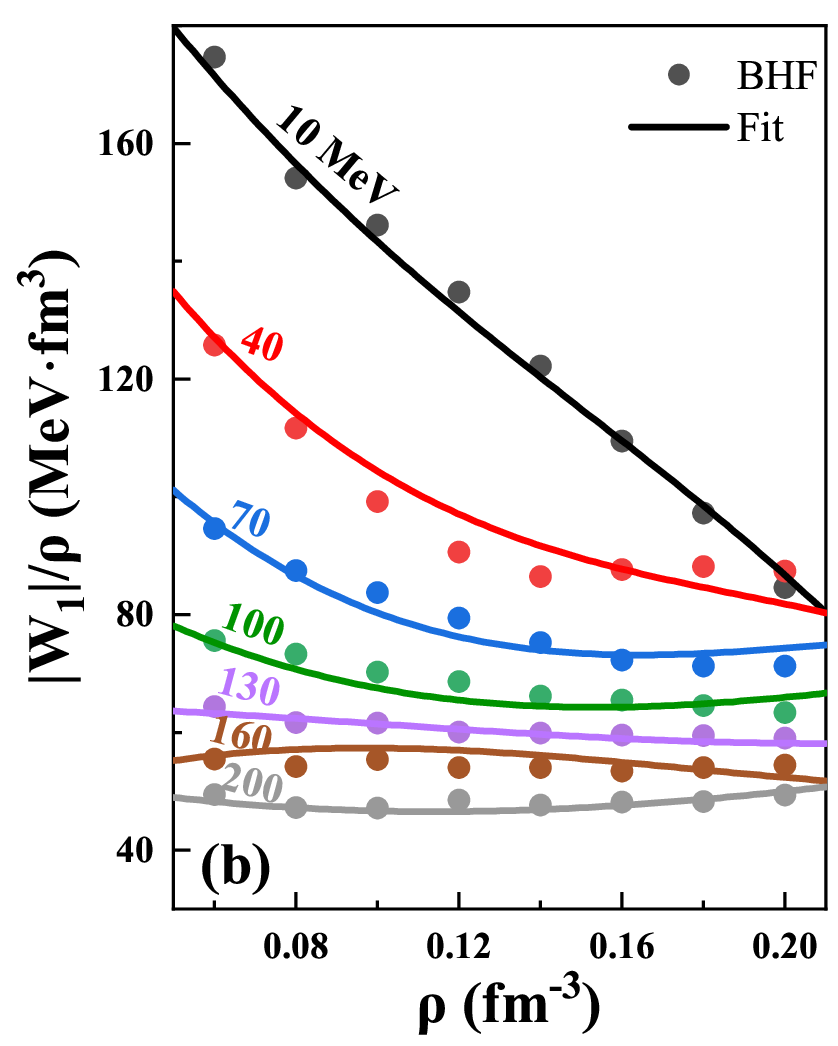}}
	\end{minipage}
	\caption{ (Color online) The solid dots represent the dependence on the density of the quantities ~$|V_1(\rho,E)|/\rho$ (a) and $|W_1(\rho,E)|/\rho$ (b) for energies $E$=10, 40, 70, 100, 130, 160, and 200 MeV by BHF theory. The Solid lines correspond to the parametrizations.
	}\label{VW1average}
\end{figure}

Fig. \ref{VW1average} shows the density dependence for the isovector part $|V_1(\rho,E)|/\rho$ and $|W_1(\rho,E)|/\rho$ of the microscopic optical potential with the energy ranging from 10 to 200 MeV, the solid dots are the results from BHF calculations, the lines are the fitted results. The real part $V_1$ can be fitted by using the expression
\begin{equation}\label{eq:V1}
	V_1(\rho,E)=\sum_{i,j=1}^4c_{ij}\rho^iE^{j-1},
\end{equation}
where $c_{ij}$ are the coefficients given in Table \ref{tab:cij-V1}. The following formula is adopted to parametrize the imaginary part $W_1$
\begin{equation}\label{eq:W1}
W_1(\rho,E)=\left[1+\frac{F}{E-\varepsilon_F}\right]^{-1}\sum_{i,j=1}^4d_{ij}\rho^iE^{j-1},
\end{equation}
where $F=1$~MeV taken from Refs.~\cite{Jeuke77,bauge98}, and $d_{ij}$ are the coefficients given in Table \ref{tab:dij-W1}.

\begin{table}[htb]
	\centering
	\caption{Coefficients ~$c_{ij}$ ~in the expression~(\ref{eq:V1})~for~$V_1(\rho,E)$~.}
	\label{tab:cij-V1}
	\setlength\tabcolsep{5mm}{
    \renewcommand{\arraystretch}{1.5}
		\begin{tabular}{ccccc}\\
			\hline\hline
			{ $c_{ij}$} & 1    & 2     & 3    & 4   \\
			\hline
		1   & $2.1445\times 10^{2}$  & $-4.5219\times 10^{0}$ & $3.5608\times 10^{-2}$  & $-9.3402\times 10^{-5}$ \\
		2   & $-2.6888\times 10^{3}$ & $9.0786\times 10^{1}$  & $-8.4421\times 10^{-1}$ & $2.3260\times 10^{-3}$  \\
		3   & $1.8634\times 10^{4}$  & $-7.0104\times 10^{2}$ & $6.7709\times 10^{0}$   & $-1.8895\times 10^{-2}$ \\
		4   & $-4.1480\times 10^{4}$ & $1.6482\times 10^{3}$  & $-1.6222\times 10^{1}$  & $4.5641\times 10^{-2}$  \\
		\hline\hline
	\end{tabular}}
\end{table}

\begin{table}[htb]
	\centering
	\caption{Coefficients ~$d_{ij}$ ~in the expression~(\ref{eq:W1})~for~$W_1(\rho,E)$~.}
	\label{tab:dij-W1}
	\setlength\tabcolsep{5mm}{
    \renewcommand{\arraystretch}{1.5}
		\begin{tabular}{ccccc}\\
			\hline\hline
			{ $d_{ij}$} & 1    & 2     & 3    & 4   \\
			\hline
		1   & $2.5909\times 10^{2}$  & $-1.5418\times 10^{0}$  & $-3.7076\times 10^{-3}$ & $3.1559\times 10^{-5}$  \\
		2   & $-1.2413\times 10^{3}$ & $-1.8332\times 10^{1}$  & $3.9026\times 10^{-1}$  & $-1.3568\times 10^{-3}$ \\
		3   & $4.1608\times 10^{3}$  & $1.5614\times 10^{2}$   & $-2.5924\times 10^{0}$  & $8.6387\times 10^{-3}$  \\
		4   & $-1.0895\times 10^{4}$ & $-1.4298\times 10^{2}$  & $3.4035\times 10^{0}$   & $-1.2178\times 10^{-2}$ \\
			\hline\hline
	\end{tabular}}
\end{table}

\begin{figure}[htb]
\centerline{\includegraphics[width=10cm]{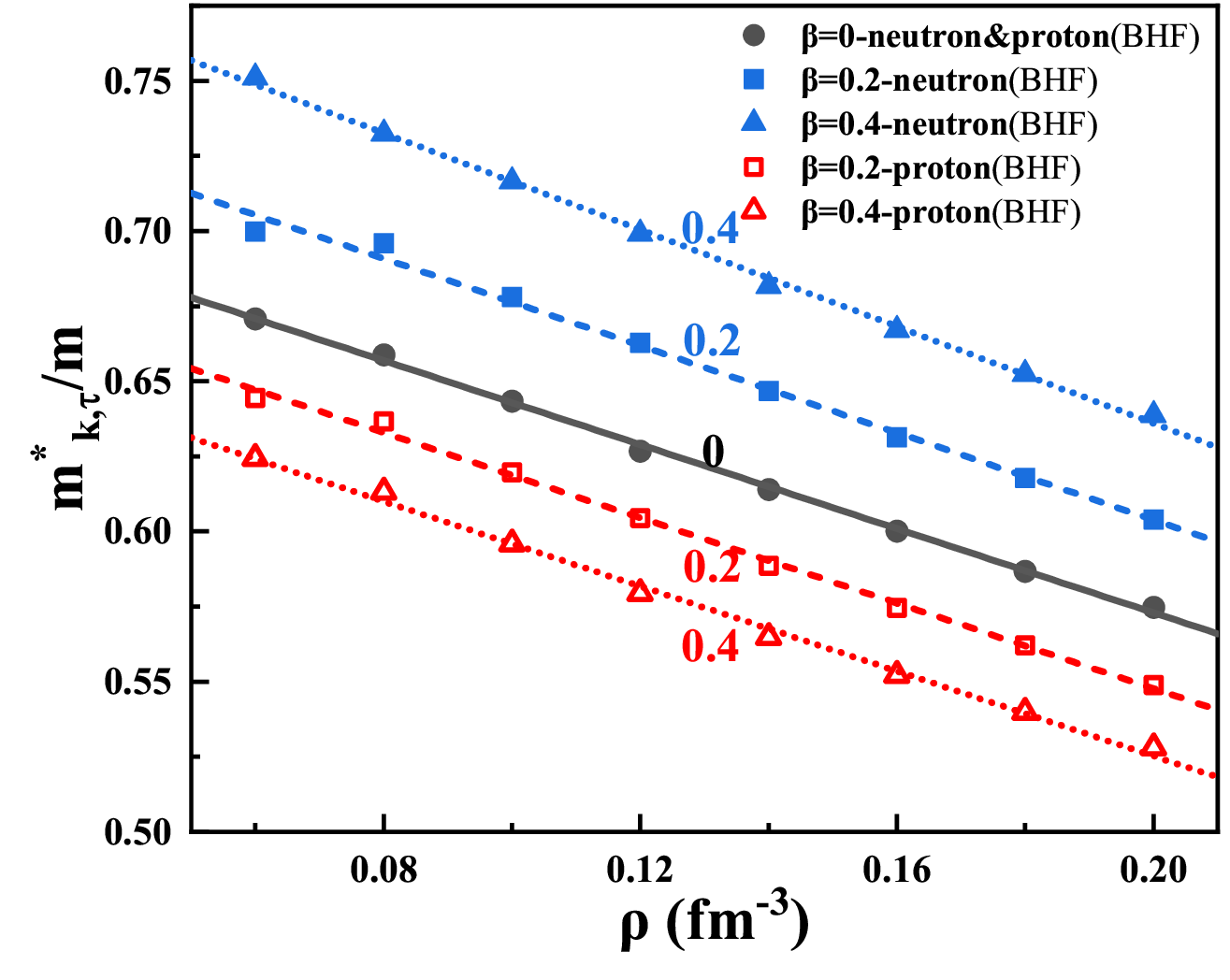}}
\caption{(Color online) The splitting of effective k-mass~$m^*_{k,\tau}/m$~of neutron and proton. The symbols and the lines are the results of BHF and fitting at $\beta$=0.0, 0.2, and 0.4, respectively.}
\label{fkmass}
\end{figure}

In Eq. (8), the k-mass $m^*_{k,\tau}/m$ applied to the imaginary optical potential serves as the following purpose: it properly accounts for the nonlocality of optical potential through momentum-dependent corrections. We use two expressions below for neutron and proton to fit the BHF effective mass, respectively,
\begin{equation}\label{eq:kmassn} \frac{m^*_{k,n}}{m}\left(\rho,\beta;k=k_F^n\right)=1-\left(\sum_{i=1}^3e_i
\beta^{i-1}\rho+\sum_{i=1}^3f_i\beta^{i-1}\right),
\end{equation}
\begin{equation}\label{eq:kmassp}	
\frac{m^*_{k,p}}{m}\left(\rho,\beta;k=k_F^p\right)=1-\left(\sum_{i=1}^3g_i
\beta^{i-1}\rho+\sum_{i=1}^3h_i\beta^{i-1}\right),
\end{equation}
where the values of coefficients $e_i$, $f_i$, $g_i$ and $h_i$ are listed in Table \ref{tab:efgh-k mass} for details. Fig. \ref{fkmass} shows the BHF effective mass of neutron and proton (the symbols) and the fitting results (the lines).
\begin{table}[htb]
	\centering
	\caption{Coefficients $~e_{i},~f_{i}~$~in the expression~(\ref{eq:kmassn})~for~${m^*_{k,n}}/{m}$~ and $~g_{i},~h_{i}~$~in the expression~(\ref{eq:kmassp})~for~${m^*_{k,p}}/{m}$~.}
	\label{tab:efgh-k mass}
	\setlength\tabcolsep{5mm}{
    \renewcommand{\arraystretch}{1.5}
		\begin{tabular}{ccccc}\\
        \hline\hline
		\multicolumn{1}{l}{} & \multicolumn{2}{c}{Neutron} & \multicolumn{2}{c}{Proton} \\
		\cmidrule(r){2-3}\cmidrule(r){4-5}
			i   & $e_i$   & $f_i$   & $g_i$   & $h_i $ \\
		\hline
			1    & $6.9950\times 10^{-1}$   & $2.8720\times 10^{-1}$   & $6.9950\times 10^{-1}$   & $2.8720\times 10^{-1}$  \\
			2    & $-9.0000\times 10^{-3}$  & $-1.4950\times 10^{-1}$  & $9.0200\times 10^{-2}$   & $1.1520\times 10^{-1}$  \\
			3    & $6.8250\times 10^{-1}$   & $-1.5250\times 10^{-1}$  & $-1.8370\times 10^{-1}$  & $1.3000\times 10^{-3}$  \\
        \hline\hline
	   \end{tabular}}
\end{table}

\subsection{MOP for neutron/proton~+~$^{40,48}$Ca scattering}

The MOP in nuclear matter discussed above cannot be directly applied to finite nuclei. The local density approximation~\cite{Jeuke77} is usually adopted to provide the connection between nuclear matter and finite nuclear optical potential. LDA ascribes to the MOP at density $\rho(r)$ the
same value as in nuclear matter with the same
value of density, with the same asymmetry parameter, and at the same energy,
\begin{equation}
	U_{LDA}(r,E)=U_{NM}(\rho(r),E,\beta(r)),
\end{equation}
where $U_{NM}$ is the optical potential in nuclear matter, and $U_{LDA}$ is the MOP in a finite nucleus. When applied to nucleon-nucleus optical potentials, LDA is known to  underestimate the root mean square radii and the surface diffuseness, being a zero-range approximation. In order to cure this weakness, we extend the LDA local density by including the effect of the finite-range interaction~\cite{Jeuke77,bauge98}. This is done by introducing a Gaussian form factor into the MOP, i.e.
\begin{equation}\label{eq:ILDA} U_{ILDA}(r,E)=(t\sqrt{\pi})^{-3}\int{U_{LDA}(r',E)}
\mathrm{exp}(-|\boldsymbol{r}-\boldsymbol{r'}|^2/t^2)d^3\boldsymbol{r'},
\end{equation}
where parameter $t$ is the width of the Gaussian. In previous works~\cite{Whitehead19,Qin24,xu12,xu16,Jeuke77,bauge98}, this parameter is usually treated as a free parameter and fitted to experimental data. Different groups adopt different values for $t$ based on the interactions they employed, it can vary from 1.1 to 1.45 fm. In the present work, $t$ = 1.4 fm is adopted in the calculations.

To evaluate our microscopic optical potential
using the ILDA, we take the n/p~+~$^{40,48}$Ca elastic scattering as an example. When applying ILDA, it is necessary to provide first the nuclear density distribution of $^{40,48}$Ca.  Negele's empirical formula\cite{negele1970} offers a useful prescription for stable nuclei, its oversimplified form inadequately describes density profiles of unstable nuclei. To be consistent with the BHF results in this work, the nuclear density
distribution is calculated within the Hartree-Fock (HF) approximation using Skyrme parametrization LNS5~\cite{Gamba11}, which is built by fitting the potential energies of symmetric and asymmetric nuclear matter given by BHF calculations.

\begin{figure}[htb]
	\begin{minipage}{0.95\linewidth}
		\centerline{\includegraphics[width=17.0cm]{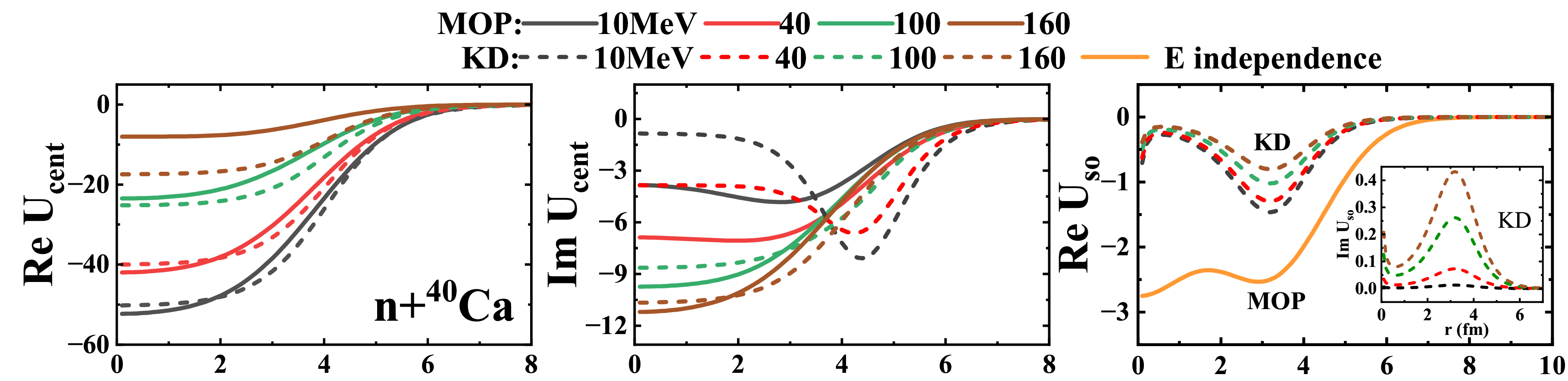}}
	\end{minipage}
	\begin{minipage}{0.95\linewidth}
		\centerline{\includegraphics[width=17.0cm]{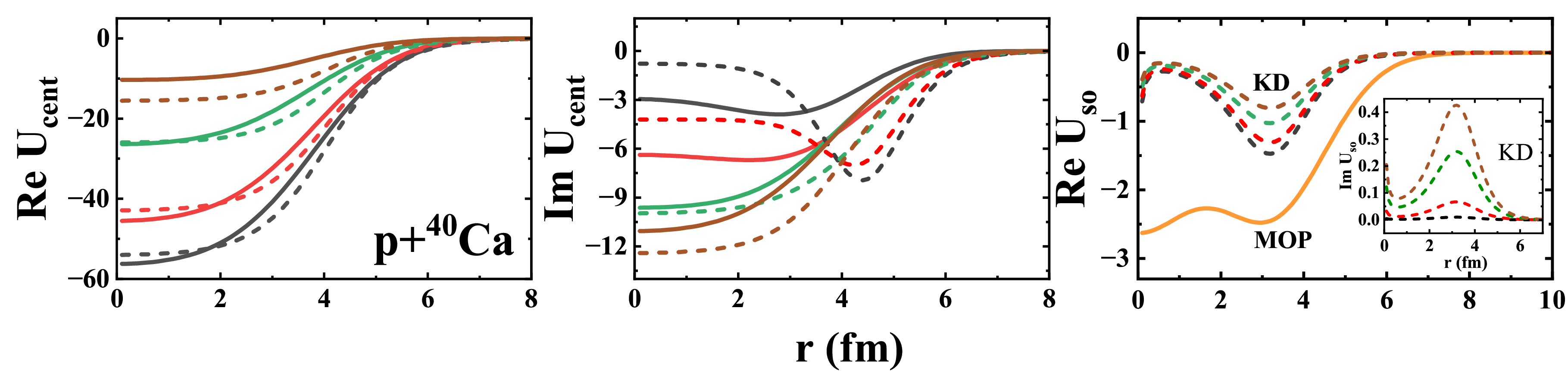}}
	\end{minipage}
	\begin{minipage}{0.95\linewidth}
		\centerline{\includegraphics[width=17.0cm]{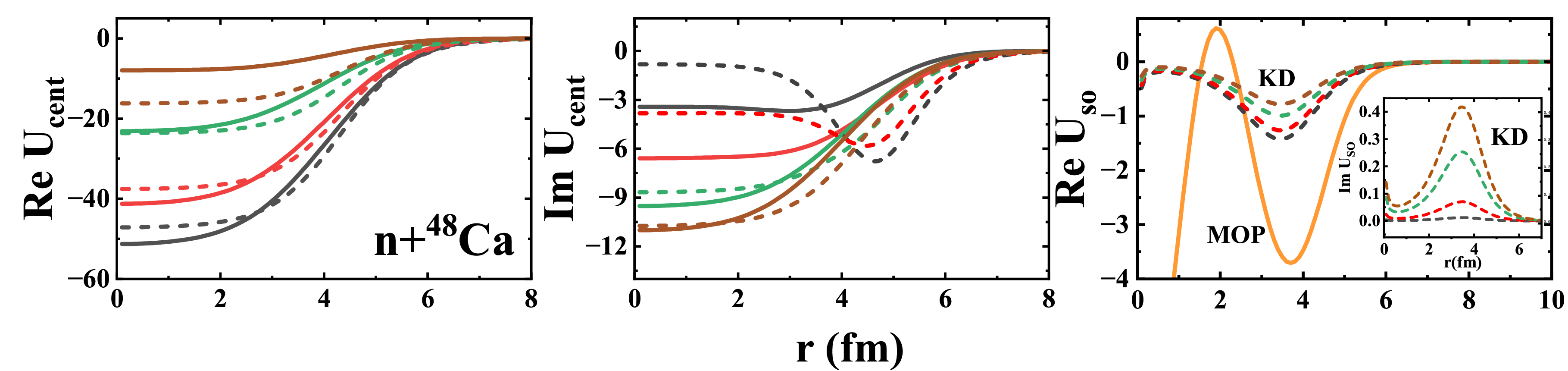}}
	\end{minipage}
	\begin{minipage}{0.95\linewidth}
		\centerline{\includegraphics[width=17.0cm]{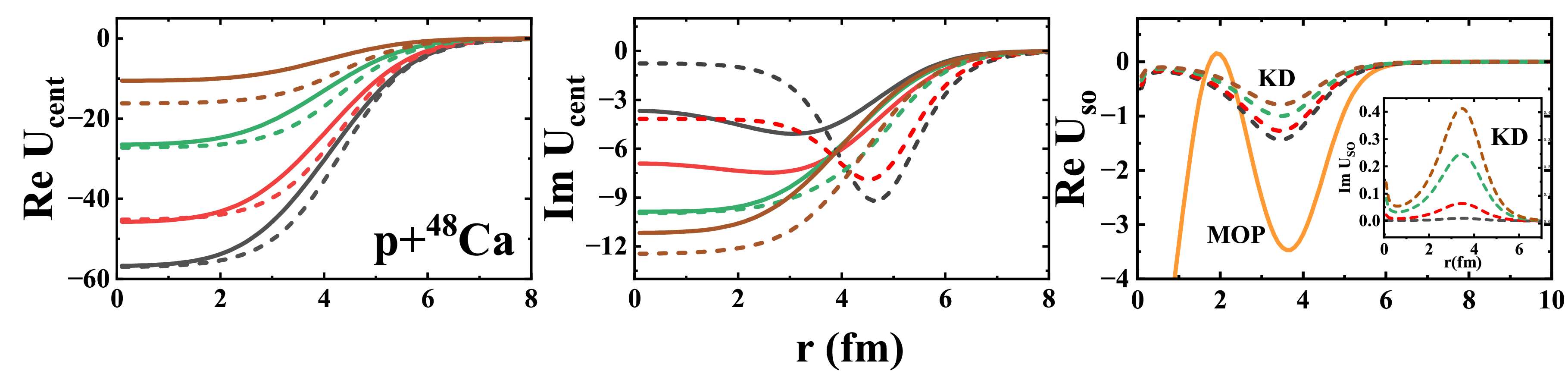}}
	\end{minipage}
	\caption{(Color online) The real (left) and imaginary (middle) central parts as well as spin-orbit part (right) of MOP as function of the radial coordinate for n/p+~$^{40,48}\mathrm{Ca}$ with the $E$ ranging from 10 to 160 MeV. Solid and dashed curves represent the results from MOP and KD potentials, respectively.}
    \label{np+Ca}
\end{figure}

In Fig. \ref{np+Ca}, we show the real central part (left) and imaginary central part (middle), and spin-orbit part (right) of the MOP as function of the radial coordinate for n/p~+~$^{40,48}$Ca with the $E$ ranging from 10 to 160 MeV, comparisons with the corresponding phenomenological global KD potentials~\cite{Koning03} are also presented in the figures. The solid and dashed lines are the results of MOP and KD potentials, respectively. In nuclear matter the spin-orbit contribution is vanishing. Using the spin-orbit potential from a Skyrme interaction, which fits simultaneously experimental data and BHF equation of state\cite{Gamba11}, is presently the best that can be done with a microscopic approach of nuclear matter. Therefore we start from the HF
calculations for finite nuclei $^{40,48}$Ca using LNS5 interaction to get the spin-orbit terms. In the HF calculation of the
spherical nucleus, we only have the real part of spin-orbit potential. The imaginary part of the spin-orbit potential below
100 MeV is usually very small~\cite{Shen09,Whitehead19,Qin24,xu12}, which is often omitted. Its contribution to MOP is also omitted in present work.

In particular, the real central part of the microscopic optical potential from LDA calculations has
a much smaller surface diffuseness compared to phenomenology.
When the improved local density approximation is employed, the comparison
to KD potential is much improved. From the left panels of Fig. \ref{np+Ca}, one can see that the overall
shapes of the real central part of the microscopic optical potentials are in
very good agreement with the global KD phenomenological
optical potentials for $E$~=~10, 40, 100 MeV, not only for isospin symmetric nucleus $^{40}$Ca, but also for isospin asymmetric nucleus $^{48}$Ca. For $E$~=~160 MeV, the depths of MOP are smaller than that of KD potentials.
The microscopic potentials of imaginary part exhibit large qualitative differences compared to KD potentials. In the middle panels of Fig. \ref{np+Ca}, pronounced surface peaks are found in KD potentials at low projectile energies 10 and 40 MeV, while the peaks in MOP are very small. For the higher energies, $E$~=~100 and 160 MeV, the shapes and the depths of imaginary part of MOP are in good agreement with the phenomenological global KD potentials for n~+~$^{40,48}\mathrm{Ca}$. For p~+~$^{40,48}\mathrm{Ca}$, the depths of imaginary part of MOP are less than that of phenomenological potentials, especially for $E$~=~160 MeV. In fact, the behavior mentioned above is a common feature in microscopic optical
potentials computed in nuclear matter approaches~\cite{Whitehead19,Petler85,Kohno84}. In the right panels of Fig. \ref{np+Ca}, the microscopic real spin-orbit potentials are compared to the phenomenological global KD potentials. As is known, our real spin-orbit potentials are energy independent, the KD spin-orbit potentials show its energy dependence. The radial shapes of the microscopic spin-orbit potentials
are found to be very similar to that of the KD potentials for $^{40}\mathrm{Ca}$. While, the differences between microscopic and phenomenological real spin-orbit potentials are found for $^{48}\mathrm{Ca}$, strong oscillation of microscopic potentials is found at the inner part of nucleus. Overall, the microscopic potential
is much deeper. The lines in the insert are the KD imaginary spin-orbit optical potentials, the strengths are relative smaller compared to the real parts.

\subsection{Describing the Scattering Observables by the present MOP}

In this subsection, the MOP constructed
in present work is evaluated through the predictions of experimental
observables of n/p~+~$^{40,48}\mathrm{Ca}$ scattering with projectile energy
below 200 MeV. We have calculated the elastic scattering differential cross
sections $d\sigma/d\Omega$, the analyzing power $A_y$ and the total/reaction cross sections using both our MOP and phenomenological global KD potential. The calculations are done within the SWANLOP package~\cite{Arellano21}, where the corrections of relativistic effects on kinematical nature at high projectile energies are included. The calculated results are also compared to the corresponding experimental data, and the experimental data adopted in our analysis are all from the EXFOR library~\cite{exfor}, which is a comprehensive
database that gathers the nuclear reaction measurements all over the world.

\begin{figure}[htb]
\centering\includegraphics[width=2.2in]{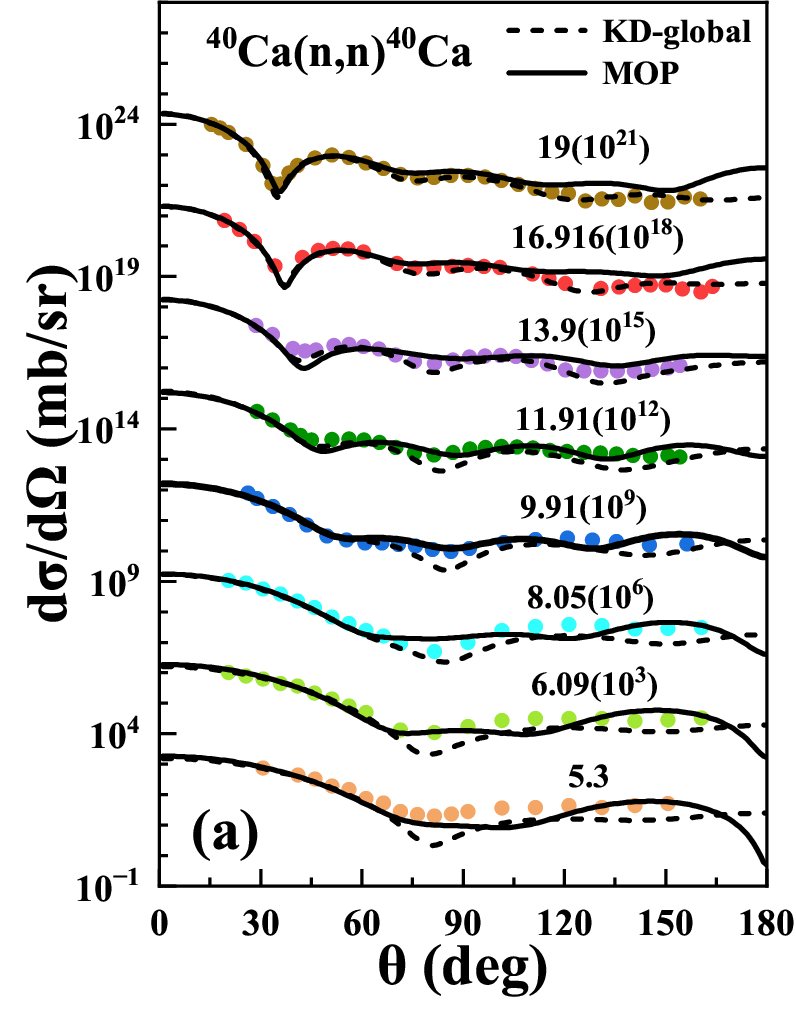}
\centering\includegraphics[width=2.2in]{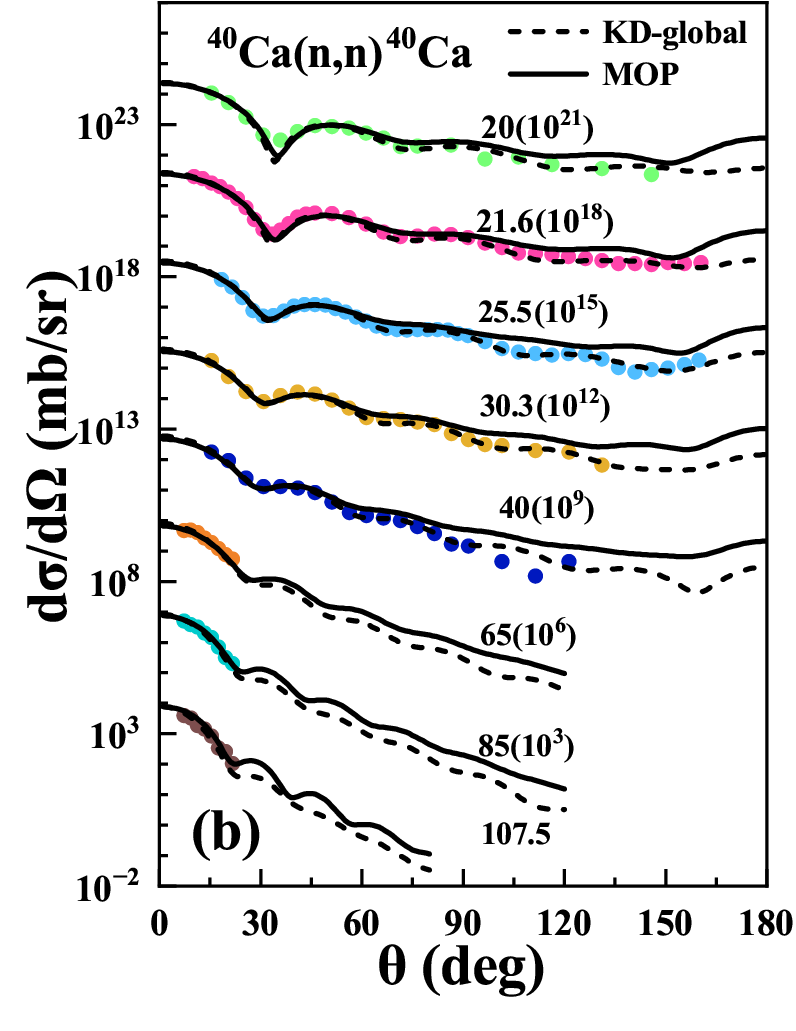}
\centering\includegraphics[width=2.2in]{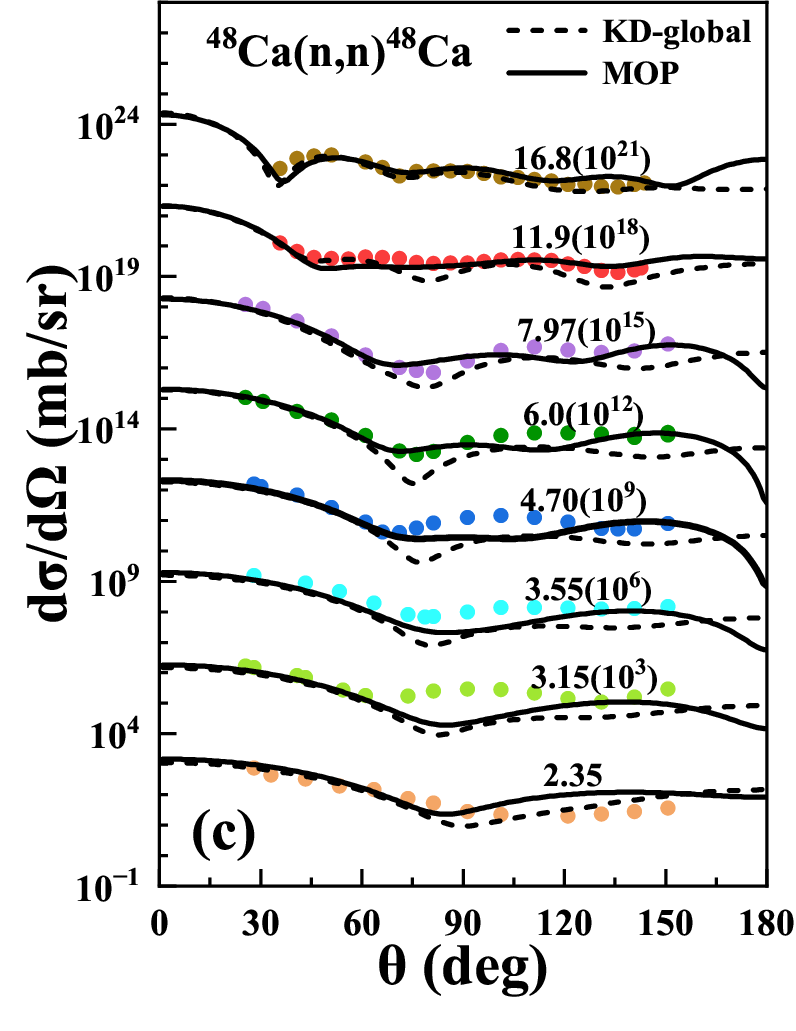}
\caption{(Color online) The predicted differential cross
sections $d\sigma/d\Omega$ for n +~$^{40,48}\mathrm{Ca}$~elastic scattering with different projectile energies by MOP (solid lines) and KD potentials (dashed lines), respectively. The corresponding experimental data (dots) are also plotted. The experimental data are taken from Ref. \cite{exfor}.  }
\label{CrsnCa40}
\end{figure}

\begin{figure}[htb]
\centering\includegraphics[width=2.2in]{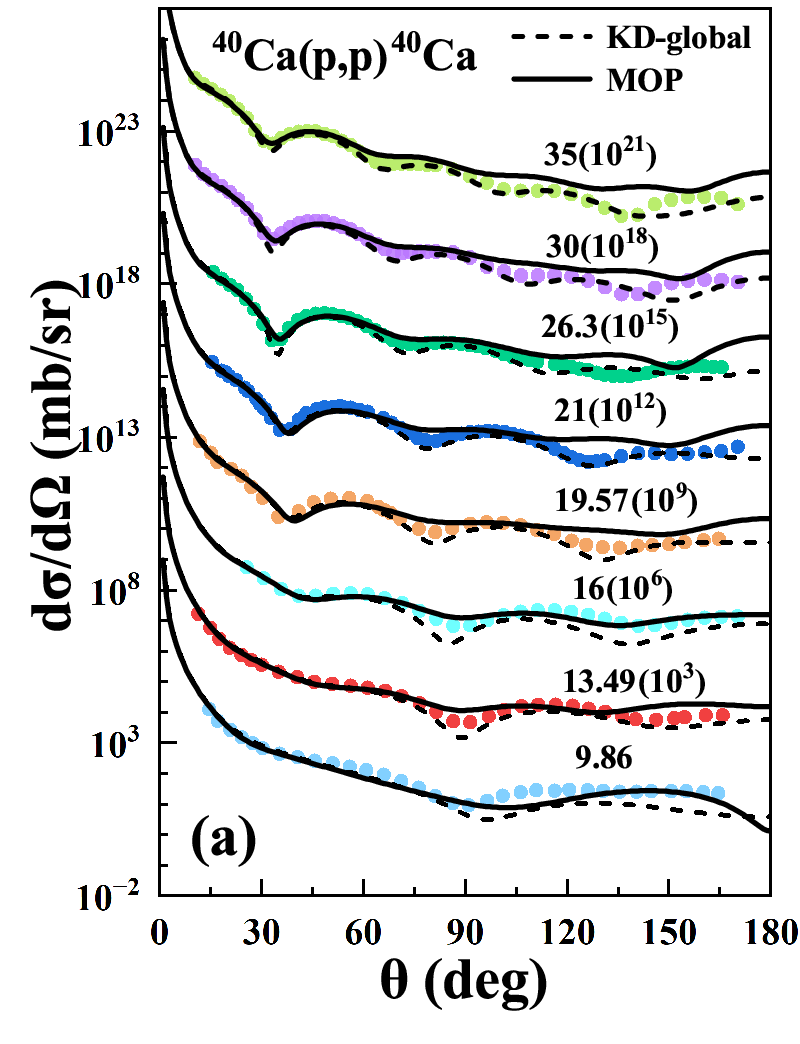}
\centering\includegraphics[width=2.2in]{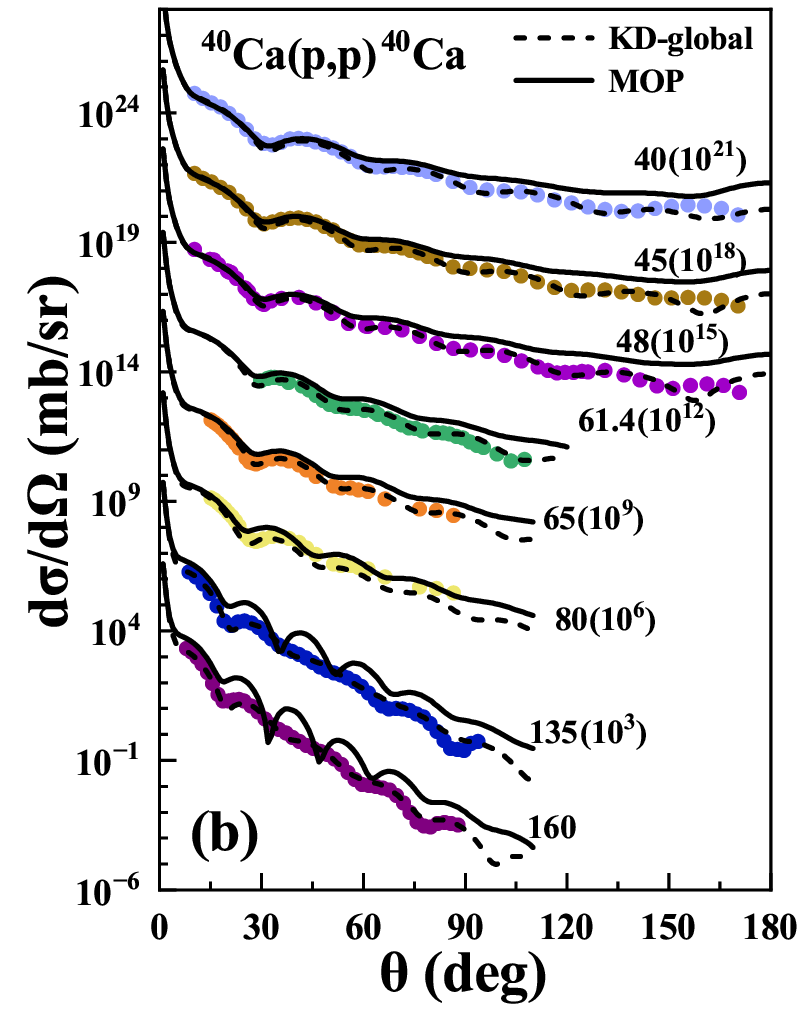}
\centering\includegraphics[width=2.2in]{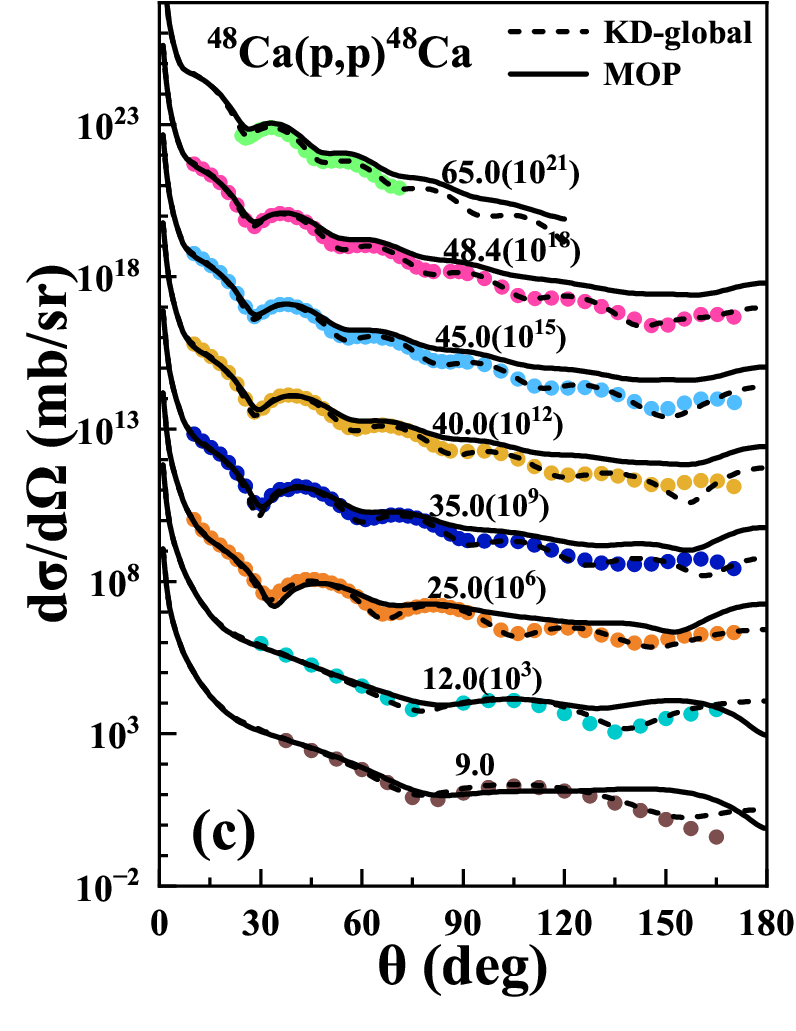}
\caption{(Color online) Similar to Fig. \ref{CrsnCa40}, but for the scattering of p +~$^{40,48}\mathrm{Ca}$.  }
\label{CrspCa40}
\end{figure}

The calculated n/p~+~$^{40,48}\mathrm{Ca}$ elastic scattering differential cross sections $d\sigma/d\Omega$ for present MOP and KD potentials with different projectile energies are shown in Fig. \ref{CrsnCa40} and \ref{CrspCa40}, respectively.
Dots with different colors are the experimental data corresponding to different projectile energies, solid and dashed lines are the results given by present MOP and KD potentials, respectively. Notice that the curves and data points at the bottom are true values, while the others are multiplied by factors of 10$^3$, 10$^6$, etc.  For the elastic scattering differential cross sections shown in Fig. \ref{CrsnCa40} and \ref{CrspCa40}, the results demonstrate that the present MOP achieves overall good agreement with experimental data across all energy ranges. For n/p~+~$^{40}\mathrm{Ca}$ results shown in (a) and (b) panels of Fig. \ref{CrsnCa40} and \ref{CrspCa40}, a detailed analysis reveals: at lower energies, the MOP and KD potential slightly underestimate the data at larger angular. In some energy regions our predictions are even better than those with the phenomenological ones, such as the differential cross sections of n~+~$^{40}$Ca at $E$~= 6.09, 9.91, and 11.91 MeV in Fig. \ref{CrsnCa40} and of p~+~$^{40}$Ca at $E$~= 9.86, 13.49, and 16.0 MeV in Fig. \ref{CrspCa40}; for the energies less than 30.3 MeV, two models exhibit comparable precision in describing the data, reproducing the angular distribution well for $\theta < 90^\circ$, particularly can capture the first minimum and subsequent peak positions; for the higher energies, the MOP and KD potentials can give better description on the data at small angle, while large-angle scattering becomes challenging for MOP, overestimates of those with larger incident energies are found. As for n~+~$^{48}\mathrm{Ca}$ results shown in (c) panel of Fig. \ref{CrsnCa40}, compared to the experimental data, it is shown that the agreement of our MOP results are much better than that of global KD results for all the energies. This may be due to that the experimental data of $^{48}\mathrm{Ca}$ are not included when fitting the global KD potential. For p~+~$^{48}\mathrm{Ca}$ results shown in (c) panel of Fig. \ref{CrspCa40}, unlike the case of neutron scattering, the global KD potentials get better agreement with the experimental data than the results given by MOP. A similar situation has also happened to $^{48}\mathrm{Ca}$, the MOP can describe the experimental data well at the small angle, while it overestimates the data at large angle. The present results show that the MOP  possesses a good predictive capability when applied to nucleon-nucleus elastic scattering differential cross sections.

\begin{figure}[htb]
\centering\includegraphics[width=2.3in]{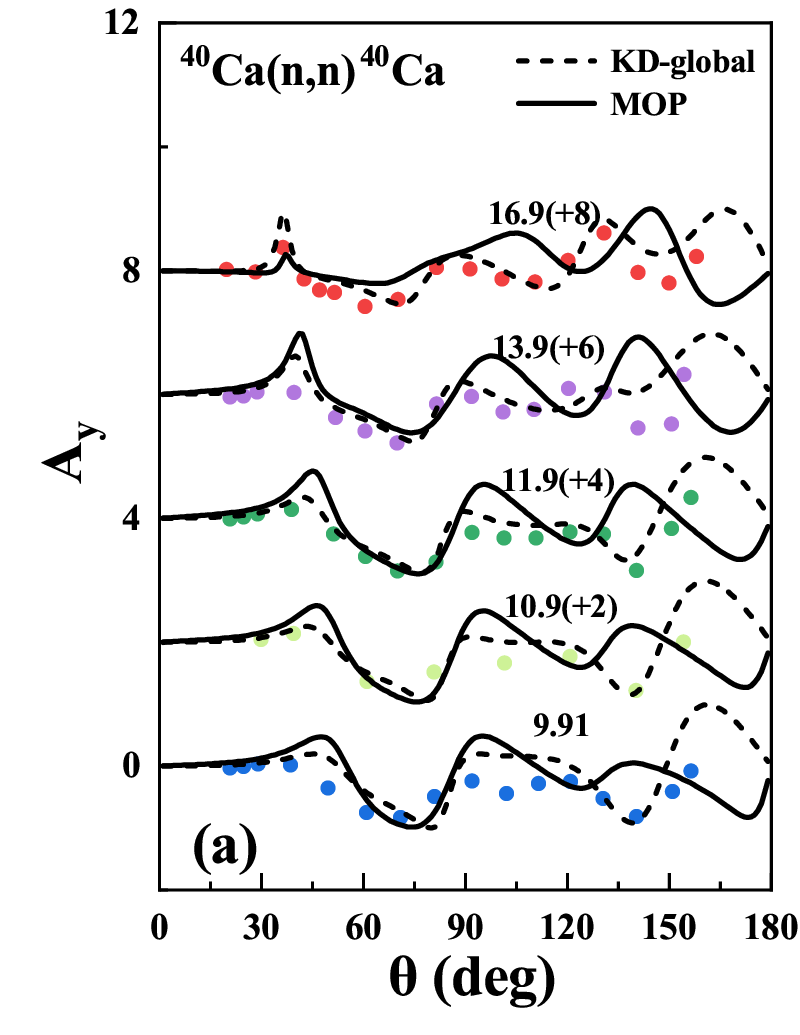}
\centering\includegraphics[width=2.3in]{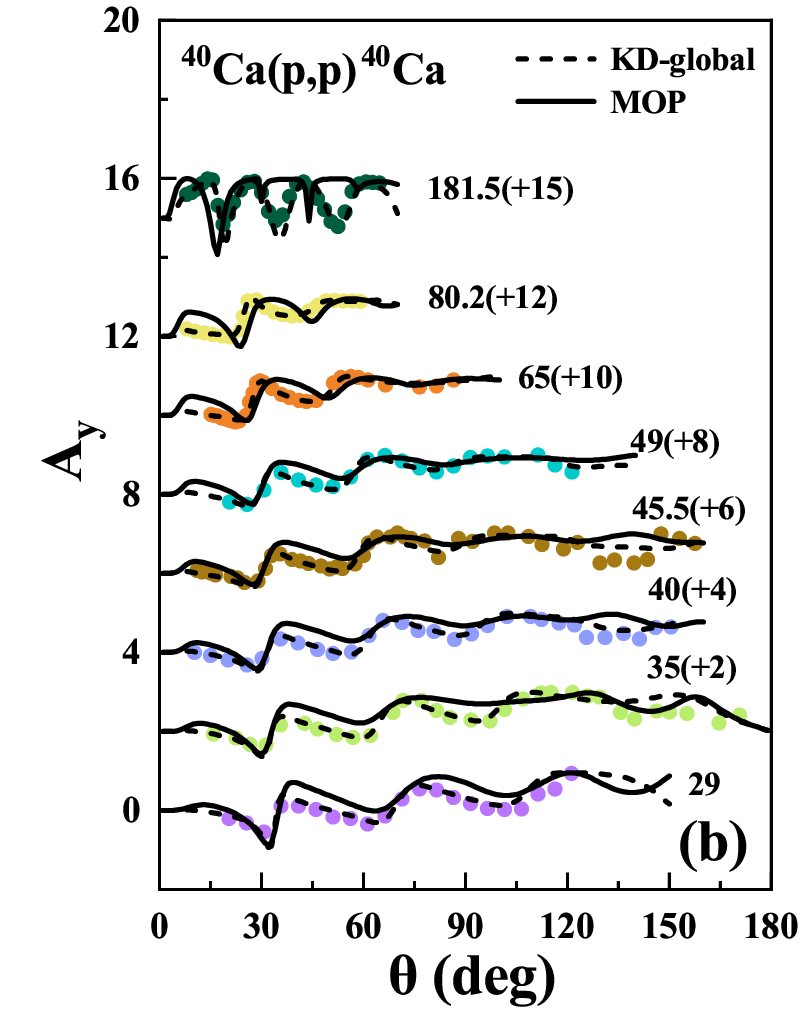}
\centering\includegraphics[width=2.2in]{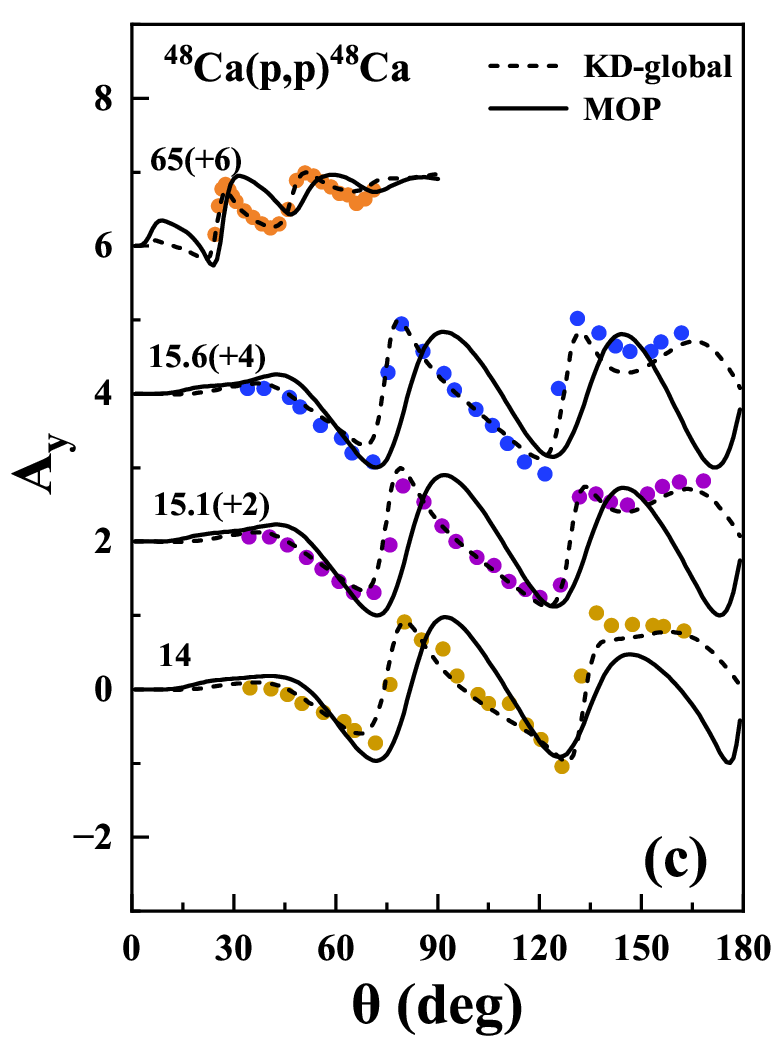}
\caption{(Color online) The predicted $A_y$ for n/p +~$^{40,48}\mathrm{Ca}$~elastic scattering with different projectile energies by MOP (solid lines) and KD potentials (dashed lines), respectively. The corresponding experimental data (dots) are also plotted. The experimental data are taken from Ref. \cite{exfor}.  }
\label{AynpCa40}
\end{figure}

Fig. \ref{AynpCa40} displays the analyzing power $A_y$ for neutrons and protons scattering off $^{40,48}$Ca targets. The solid and dashed lines are the predicted results of MOP and KD potentials, respectively. The
lines and data points at the bottom are true values, while the
others are offset by factors of +2, +4, etc. For n + $^{40}$Ca (Fig. \ref{AynpCa40}(a)), the experimental data are relatively scarce and limited to lower incident energies, whereas for p + $^{40}$Ca (Fig. \ref{AynpCa40}(b)) there are more extensive measurements.
In Fig. \ref{AynpCa40}(a), the MOP calculations reproduce the positions of $A_y$ peaks and valleys well for the angel less than 120$^\circ$, whereas the amplitudes are not ideal. For the larger angle, the predicted peaks or valleys show phase shift (either advanced or delayed) compared to the results of KD potentials and data. Fig. \ref{AynpCa40}(b) demonstrates good agreement between MOP results and experimental data for incident energies less than 100 MeV. For n + $^{48}$Ca, there is no experimental data for the analyzing power $A_y$. For p + $^{48}$Ca, we find the experimental data at four different energies as shown in Fig. \ref{AynpCa40}(c), including 14.0, 15.1, 15.6 and 65.0 MeV. For $A_y$ at the low energies, the MOP can reproduce the positions of valleys very well, but delayed peaks are found. The amplitudes of the analyzing power are comparable to the experimental data. The $A_y$ predicted by MOP for 65.0 MeV is overall delayed compared to the experimental data. It is not
surprising to find that the results from KD potentials can reproduce the experimental data very well.

\begin{figure}[htb]
	\begin{minipage}{0.49\linewidth}
		\centerline{\includegraphics[width=7.0cm]{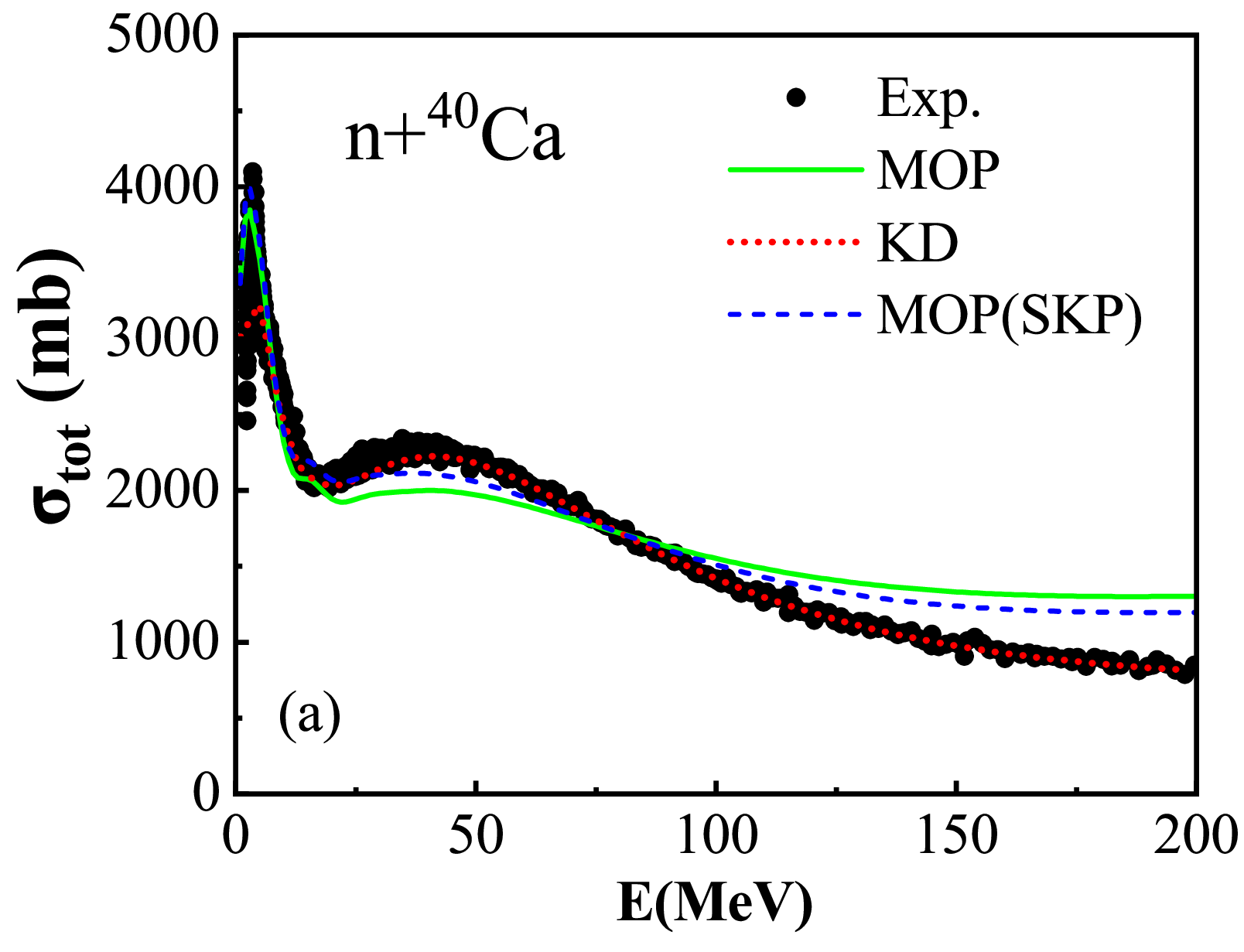}}
	\end{minipage}
	\hfill
	\begin{minipage}{0.49\linewidth}
		\centerline{\includegraphics[width=7.0cm]{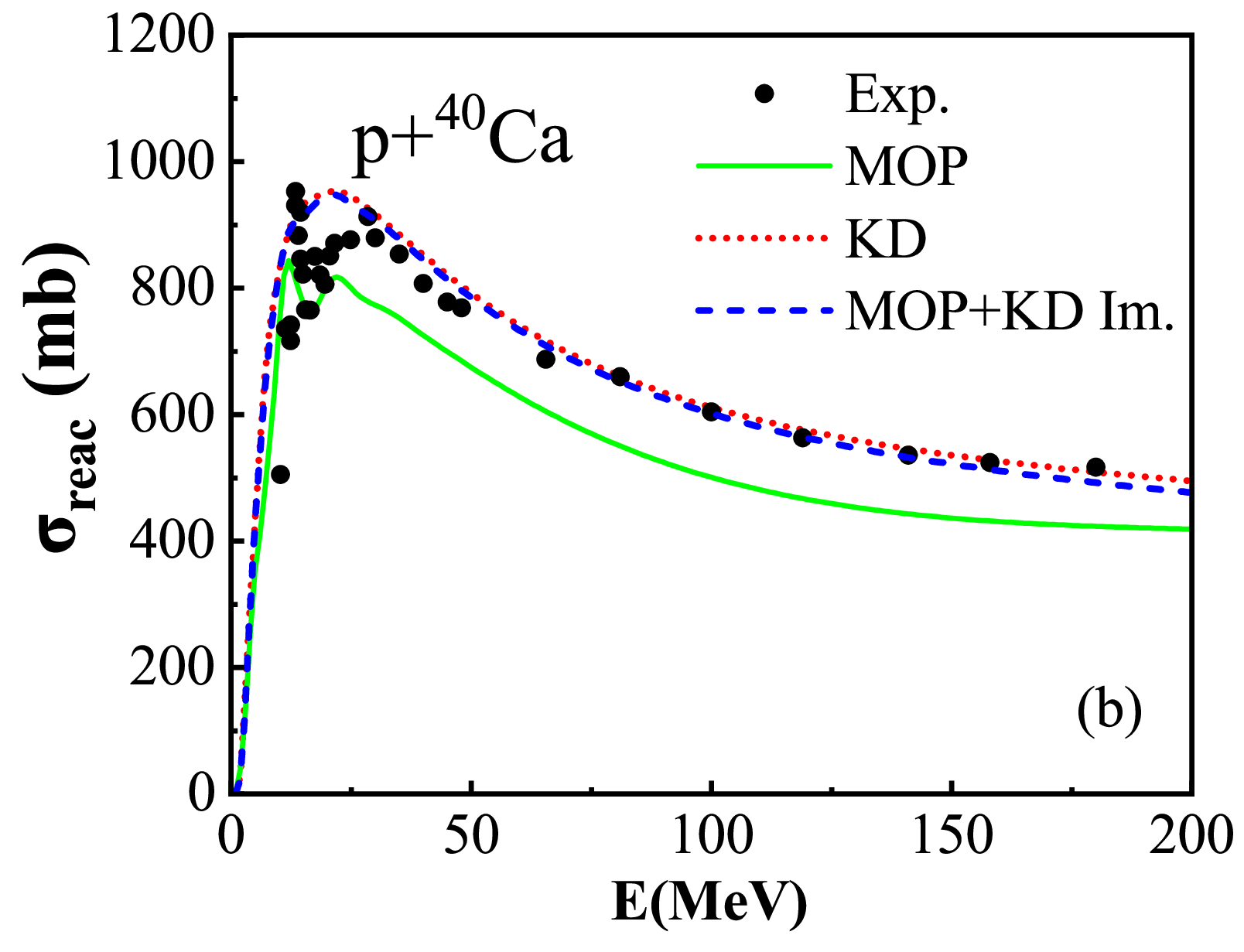}}
	\end{minipage}
	\begin{minipage}{0.49\linewidth}
		\centerline{\includegraphics[width=7.0cm]{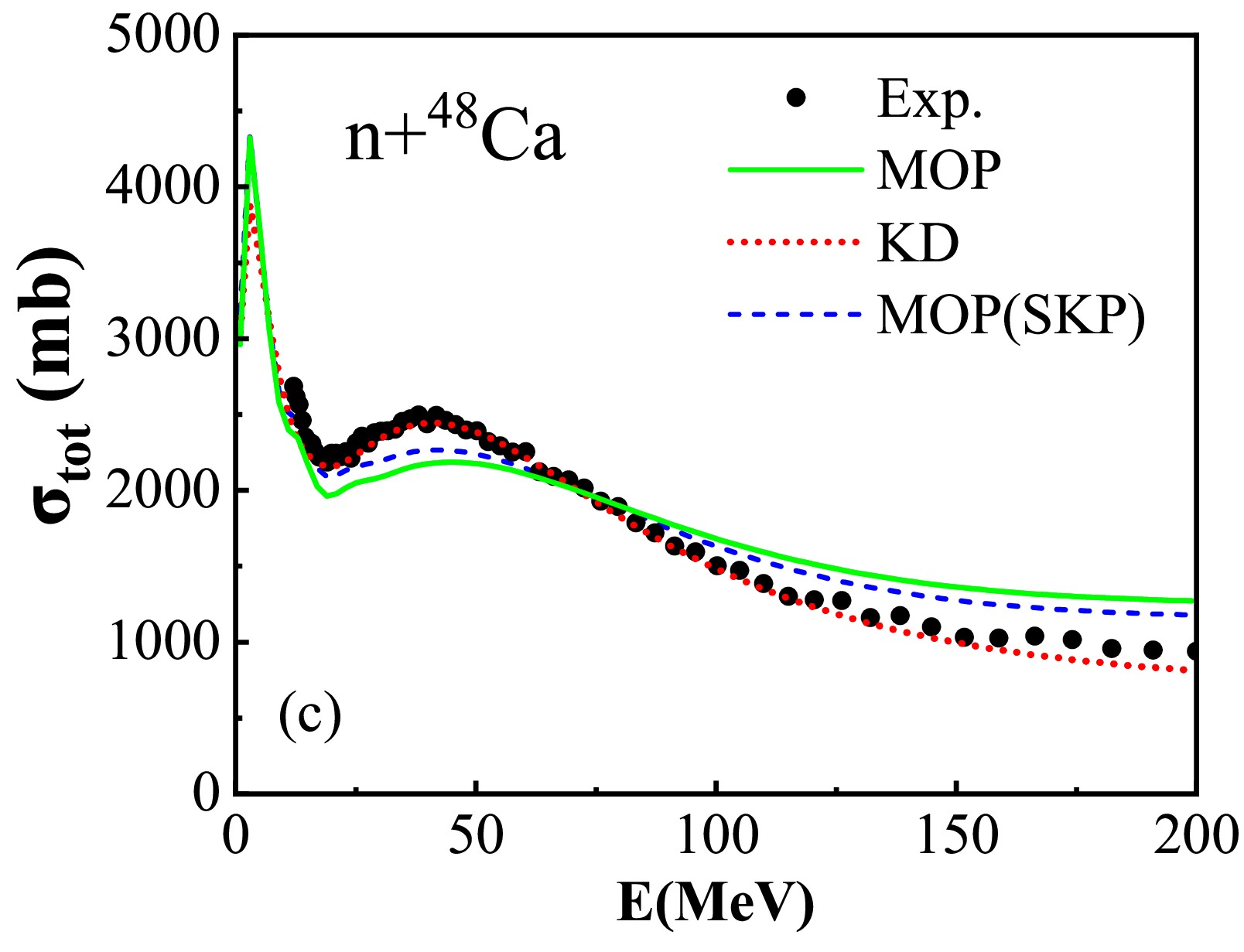}}
	\end{minipage}
	\hfill
	\begin{minipage}{0.49\linewidth}
		\centerline{\includegraphics[width=7.0cm]{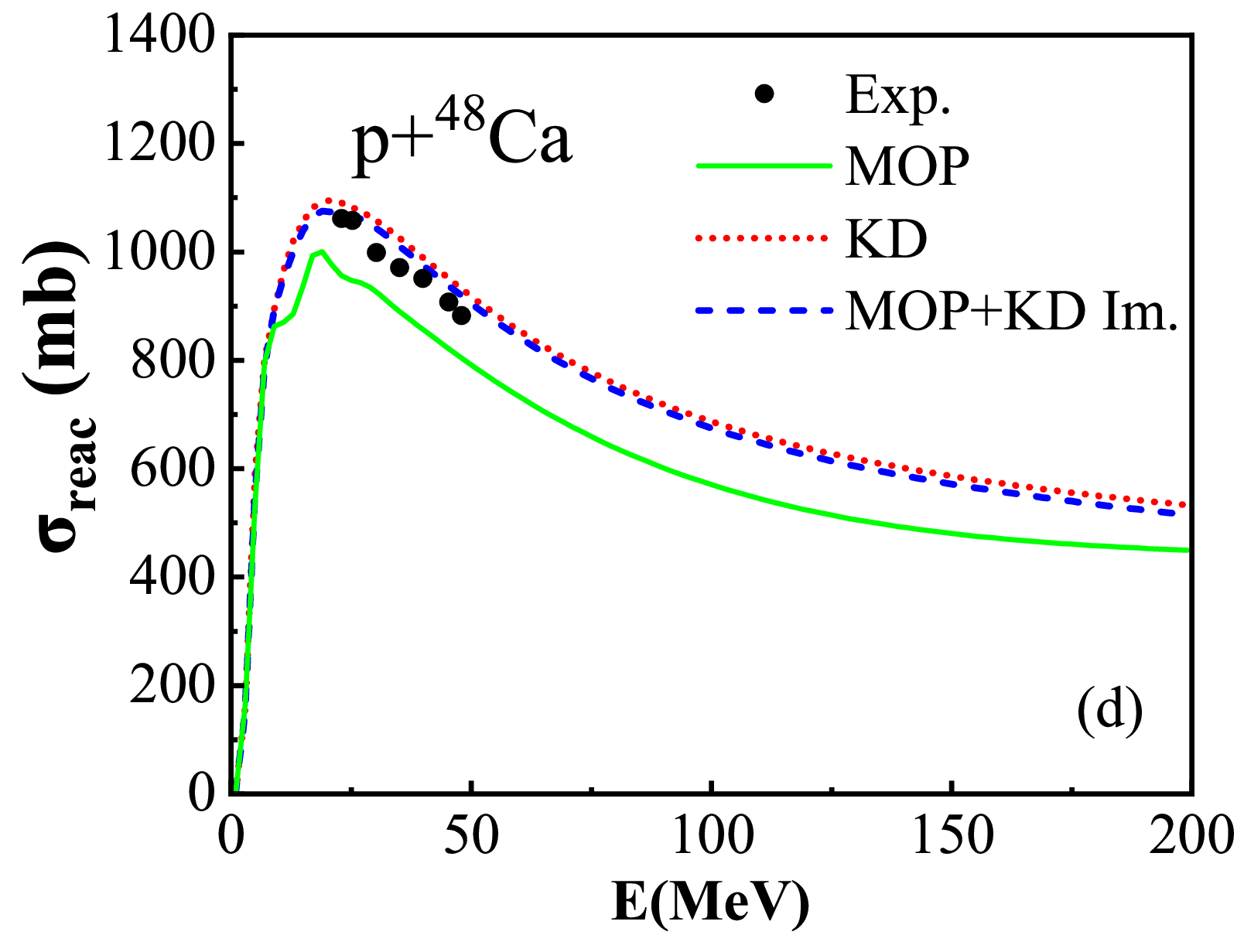}}
	\end{minipage}
	\caption{(Color online) The predicted total (total reaction) cross sections for n/p +~$^{40,48}\mathrm{Ca}$~reactions by MOP (solid lines) and KD potentials (short-dotted lines), respectively. The short-dashed lines in Fig. 8 (a) and (c) are the results obtained by using the densities given by SKP interaction. The short-dashed lines in Fig. 8 (b) and (d) are the results given by replacing the MOP imaginary potentials with KD imaginary potentials. The corresponding experimental data (dots) are also plotted. The experimental data are taken from Ref. \cite{exfor}.}\label{RcnpCa40}
\end{figure}

Fig. \ref{RcnpCa40} (a) and (c) show the total cross sections for neutron
scattering off $^{40,48}$Ca calculated by the MOP (solid lines) and KD potentials (short-dotted lines), respectively. For
comparison, we also show the experimental data. It is obvious that the results from KD potentials give good description on the experimental data. We see that the purely microscopic optical potentials predict similar behaviours as the experimental data both for $^{40}$Ca and $^{48}$Ca, the calculations underestimate the data for energies between 20 and 80 MeV, and overestimates the data for energies larger than 80 MeV. We find that the results can be improved if we adopt some more sophisticated Skyrme interactions when doing ILDA, such as the SKP interaction. The short-dashed lines in Fig. \ref{RcnpCa40} (a) and (c) are the results obtained by using the densities given by SKP interaction, the results get better than the one given by LNS5 interaction. In Fig. \ref{RcnpCa40} (b) and (d), we show the total reaction cross sections for proton
scattering off $^{40,48}$Ca calculated by the MOP (solid lines) and KD potentials (short-dotted lines), respectively. The dots are the experimental data. For $^{40}$Ca, the experimental data cover a wide range of energies, from 10 to 180 MeV. For $^{48}$Ca, there is only a few experimental data in between of 22 to 48 MeV. The KD potentials give good agreement with the experimental data both for $^{40}$Ca and $^{48}$Ca in Fig. \ref{RcnpCa40} (b) and (d). We see that the MOP predicts good
total reaction cross sections for energies less than 25 MeV for $^{40}$Ca, for
energies larger than 25 MeV, the total reaction cross sections are underestimated
by our microscopic potential. The short-dashed line in Fig. 8 (b) is the result given by replacing the MOP imaginary part with KD imaginary potential. As a whole, the replacement
of the phenomenological imaginary part leads to a
significant improvement in the description of the total reaction
cross section of $^{40}$Ca. For $^{48}$Ca, as shown in Fig. \ref{RcnpCa40} (d), the MOP give similar cross sections to KD potential at energies less than 10 MeV, and the predicted cross sections by MOP are less than the data and the values given by KD potential at energies larger than 10 MeV. Again if we replace the MOP imaginary part with KD imaginary potential for $^{48}$Ca, the results will be much better.

\section{Summary and Perspectives}\label{sec4}

In this work, we study the microscopic optical potential for finite nuclei in the framework of BHF theory. The nucleon self-energies at different densities and asymmetry parameters are obtained by solving the Brueckner-Bethe-Goldstone equation with Argonne V18 realistic NN interaction and the microscopic TBF. The BHF results are parametrized with a set of analytic forms, which permit to handle with ease the optical potentials in nuclear matter. The ILDA intimately relates the density and isospin dependence of MOP in nuclear matter with the
radial dependence in finite nuclei. The density distribution of a finite nucleus is calculated from the HF theory with the LNS5  interaction, which is the Skyrme-like interaction obtained by fitting the results of BHF calculations. The spin-orbit potentials are obtained within the HF approximation for finite nuclei $^{40,48}$Ca using LNS5 interaction. The free parameter $t$, which characterizes the ILDA, is fixed to be $t$ = 1.4 fm for all calculations in present work.

As an example of applications of our MOP, we investigate the elastic
scattering of n/p + $^{40,48}$Ca with incident energies below 200 MeV. The
predicted observables, such as the elastic scattering differential cross
sections, analyzing powers and total/reaction cross sections, are compared with
experimental data and with results from phenomenological global
KD potentials. Overall, the predictions of MOP are found to be consistent with the experimental
data for $^{40,48}$Ca targets. The results of the microscopic optical potential is comparable to the widely
used phenomenological global KD optical potentials. To further evaluate the
performance of MOP, we will extent the targets to a wide range of mass numbers $56\leq A \leq 208$ with
incident energies up to 200 MeV, and systematically apply the MOP to nucleon - nucleus scattering observables. This will be
presented in a forthcoming paper.
As shown in the text, MOP describes the scattering processes rather
well in most of the cases. However, MOP still has the room to be improved. For example, the MOP imaginary parts at low energy region are different from the phenomenological global KD potentials. Since in the present BHF calculations only the first order of mass operators is considered to contribute to the nucleon self-energy, we must take into account  higher order contributions~\cite{Whitehead19}. In the present MOP only the real part of the rearrangement term is considered, whereas the imaginary part that improves the imaginary part of MOP is expected to play an important role in describing the reaction cross section. In the future we
plan to compute the spin-orbit potential from the improved density matrix expansion method~\cite{Bogner09,Gebre10} instead of the HF spin-orbit potential used in MOP. We will also explore a wider range of realistic nucleon-nucleon interactions
in order to provide a more comprehensive estimate of the
theoretical uncertainties~\cite{King19,Baker22,Kyle25}. Finally, we propose to address the problem of extracting from the new scattering data information on the symmetry energy in the region around the saturation density~\cite{symme}.

\section{ACKNOWLEDGEMENTS}

 L. G. Cao thanks Zhongyu Ma and Gianluca Col\`o for helpful discussions. This work is partly supported by the National Natural Science Foundation of China under Grant Nos. 12275025, 12375117, 11975096, 12135004, 11961141004 and the Fundamental Research Funds for the Central Universities under Grant Nos. 2020NTST06.

\end{document}